\documentclass[12pt]{article}
\usepackage{a4,epsfig}
\usepackage{lineno}
\newcommand\eps{\varepsilon}

\begin{document}
\title{Strap-on magnets: a framework for rapid prototyping of magnets and
  beam lines}
\author{V. Ziemann, Uppsala University}
\date{\today}
\maketitle
\begin{abstract}\noindent
  We describe a framework to assemble permanent-magnet cubes in 3D-printed
  frames to construct dipole, quadrupole, and solenoid magnets, whose field,
  in the absence of iron, can be calculated analytically in three spatial
  dimensions. Rotating closely spaced dipoles and quadrupoles in opposite
  directions allows us to adjust the integrated strength of a multipole.
  Contributions of unwanted harmonics are calculated and found to be moderate.
  We then combine multiple magnets to construct beam-line modules: 
  chicane, triplet cell, and solenoid focusing system. 
\end{abstract}
%
%
\section{Introduction}
Various types of magnets play important roles in many physics laboratories.
Unfortunately, they are often expensive and require a long lead times to
order. Here we propose a complementary method: we construct multipole
magnets from standard-size permanent-magnet cubes, which are reasonably
inexpensive and easy to obtain. Moreover, we suggest to hold them in 3D-printed
frames or CNC-machined aluminum frames that can be produced in most small
workshops. By constructing the frames in two halves they can be wrapped-around
or strapped-on to circular pipes and easily fixed with non-magnetic screws,
which makes retrofitting them to existing beam lines very simple.
Figure~\ref{fig:frames} illustrates the frames for a dipole and a quadrupole
constructed from eight cubes, each.
\par
The orientation of the cubes in the frames follows the Halbach method~\cite{KH1},
adapted to square~\cite{VZAP} instead of trapezoidal magnet segments. In an $M$-magnet
dipole, subsequent magnets along the azimuth are rotated by $2\times 360^o/M$.
The left-hand image in Figure~\ref{fig:frames} illustrates this for a dipole
made of eight cubes. In the frame for an eight-magnet quadrupole, shown on the
right-hand image in Figure~\ref{fig:frames}, subsequent magnets rotate by
$3\times 360^o/8$ as illustrated by the little notch in each of the eight holes
for the cubes. 
\par
In order to calculate the magnetic fields from the cubes in three dimensions we
use the closed-form expressions from~\cite{VZCFE}, which allow us to prepare
MATLAB scripts to determine the relevant field quantities and the multipole
contents of the magnets in parameterized form. This proves very convenient to
adapt the design to different sized cubes, different geometries, and other
multipolarities.
\par
The organization of this report is as follows: after we discuss the design of
the frames with the OpenSCAD software, we characterize eight-magnet dipoles,
quadrupoles, and solenoids made of 10\,mm cubes. Then we combine two dipoles
or quadrupoles to produce dipoles and quadrupoles whose field strength can
be varied. In order to illustrate their use, we combine several magnets to
beam-line modules.
\section{Frames}
%
\begin{figure}[tb]
\begin{center}
\includegraphics[width=0.49\textwidth]{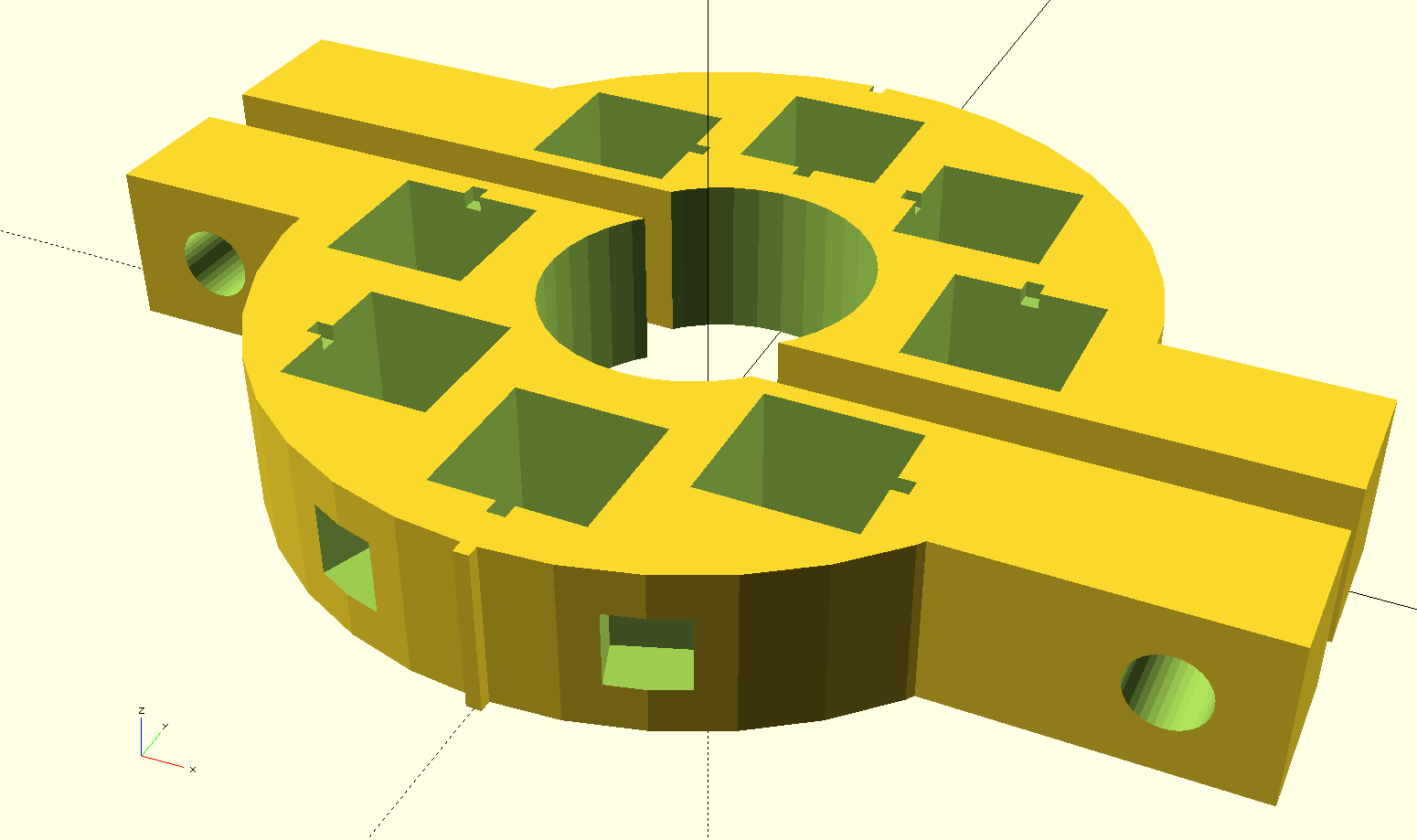}
\includegraphics[width=0.49\textwidth]{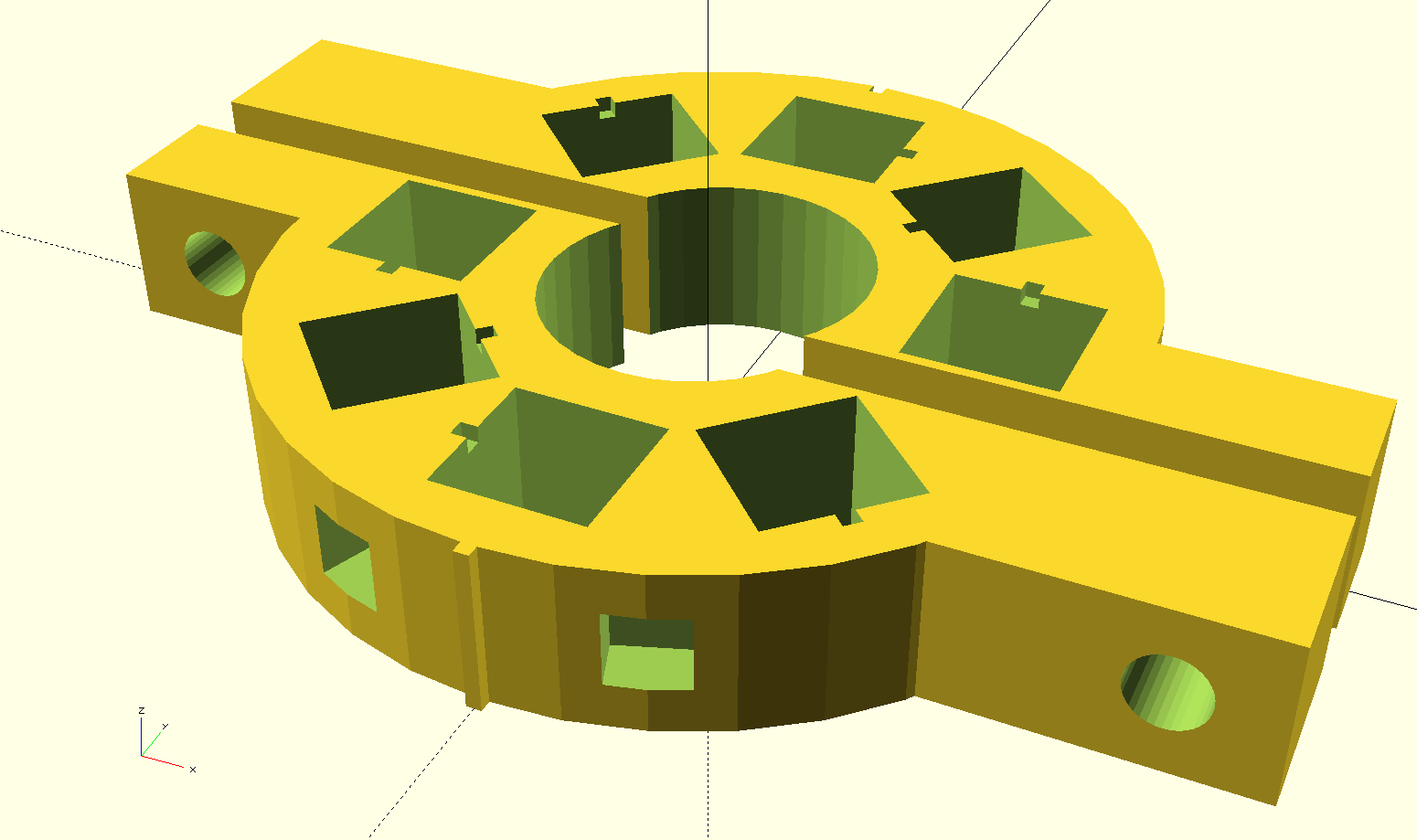}
\end{center}
\caption{\label{fig:frames}The frames for dipoles (left) and quadrupoles (right).}
\end{figure}
We use OpenSCAD to prepare solid models of the frames to hold the permanent-magnet
cubes, because this CAD software is easily scriptable, such that parameters for
different-sized magnets of multipoles only require changing a few parameters.
The frame can then be exported in a format that is compatible with 3D printers
or CNC milling machines, which allows for short prototyping cycles. 
\par
The frame consists of a cylinder with an outer radius of $r_o=28\,$mm, a 1\,mm
notch and a groove on its outer periphery to tell up from down, and 5\,mm square
holes to later attach handling rods. Furthermore there are rectangular solids to
later hold screws---the horizontal circular holes shown in Figure~\ref{fig:frames}.
Moreover a circular hole with a radius of $r_i=10\,$mm is removed from
the center. The holes for the 10\,mm cubes are then ``stamped out'' from this
base frame. Each of the holes is born at the center, receives the little notch,
and is rotated by an angle $m\phi$, where $m$ is the multipolarity of the magnets
and $\phi$ is the azimuthal position of the cube in the frame. Then it is moved
outwards by $o=18\,$mm along the $x$-axis before rotated once again by $\phi$,
which puts the hole in place. We point out that the notches on the holes are only
for orientation, in the magnets we use that direction to indicate the
``feather'' (an not the ``tip'') of the arrows that indicate the easy axis of
the magnets. Typically the square holes for the cubes are 0.1\,mm larger than
the magnet cubes themselves, which makes inserting the cubes easier. Inside the
holes they are then easily fixated with super-glue.
\par
Note that these frames have semi-circular holes with a diameter of $2r_i=20\,$mm
in their center. Again, in order to account for finite tolerances of the frame
and the pipe, normally the diameter needs to be increased a little, say by 0.1\,mm.
This makes it easy to wrap the frames around a circular pipe, align
the holes for the screws, insert the screws and tighten them. This fixes the frame
in place. Moving it along the pipe or rotating it only involves loosening the
screws, moving the frame with the magnets, and tightening the screws again. 
\par
Let us now investigate the properties of the magnetic field generated by cubes
in such frame---first out are dipoles.
\section{Dipole}
\label{sec:dipone}
%
\begin{figure}[tb]
\begin{center}
\includegraphics[width=0.47\textwidth]{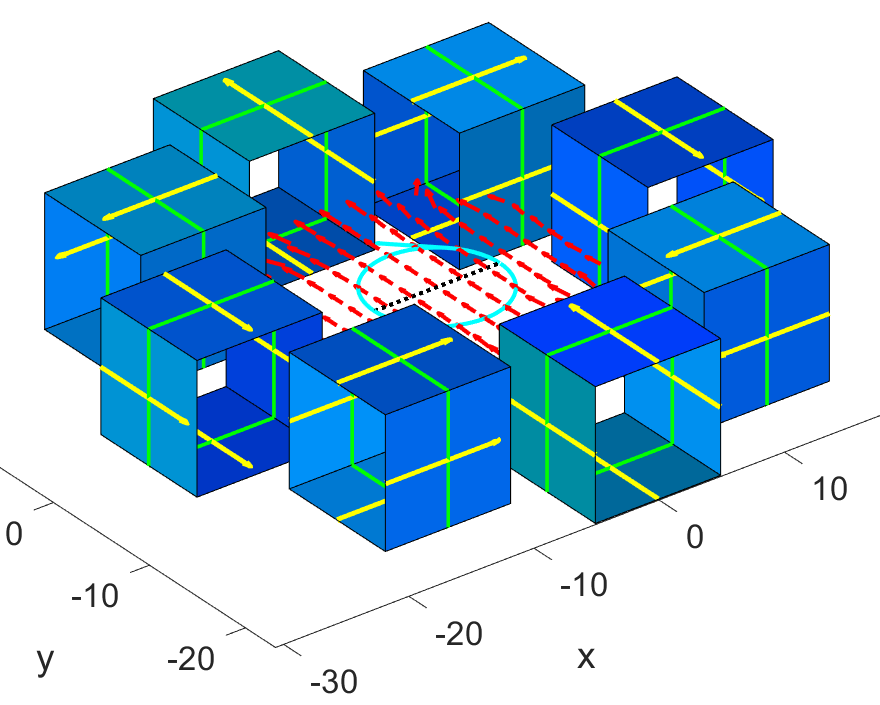}
\includegraphics[width=0.47\textwidth]{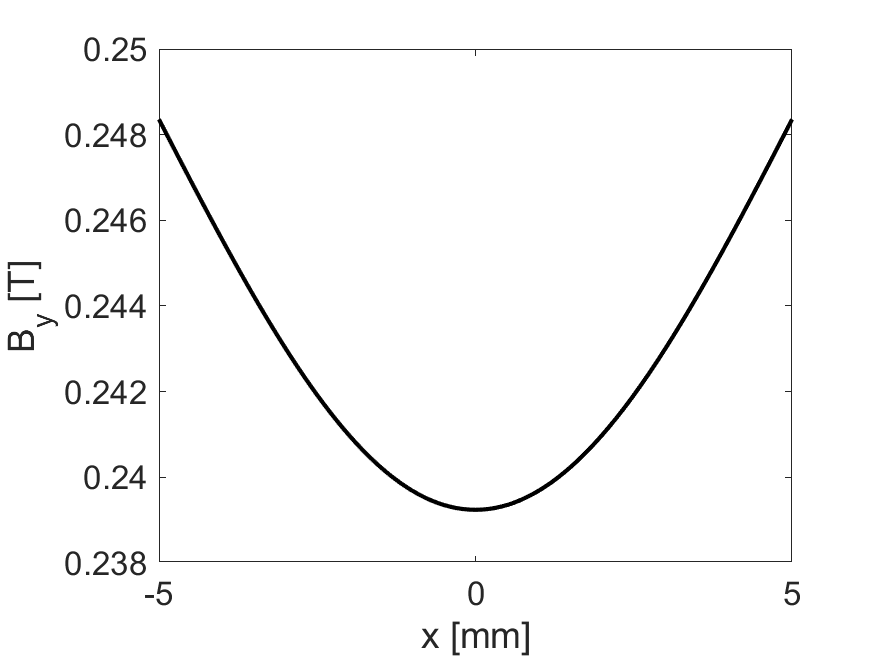}
\includegraphics[width=0.47\textwidth]{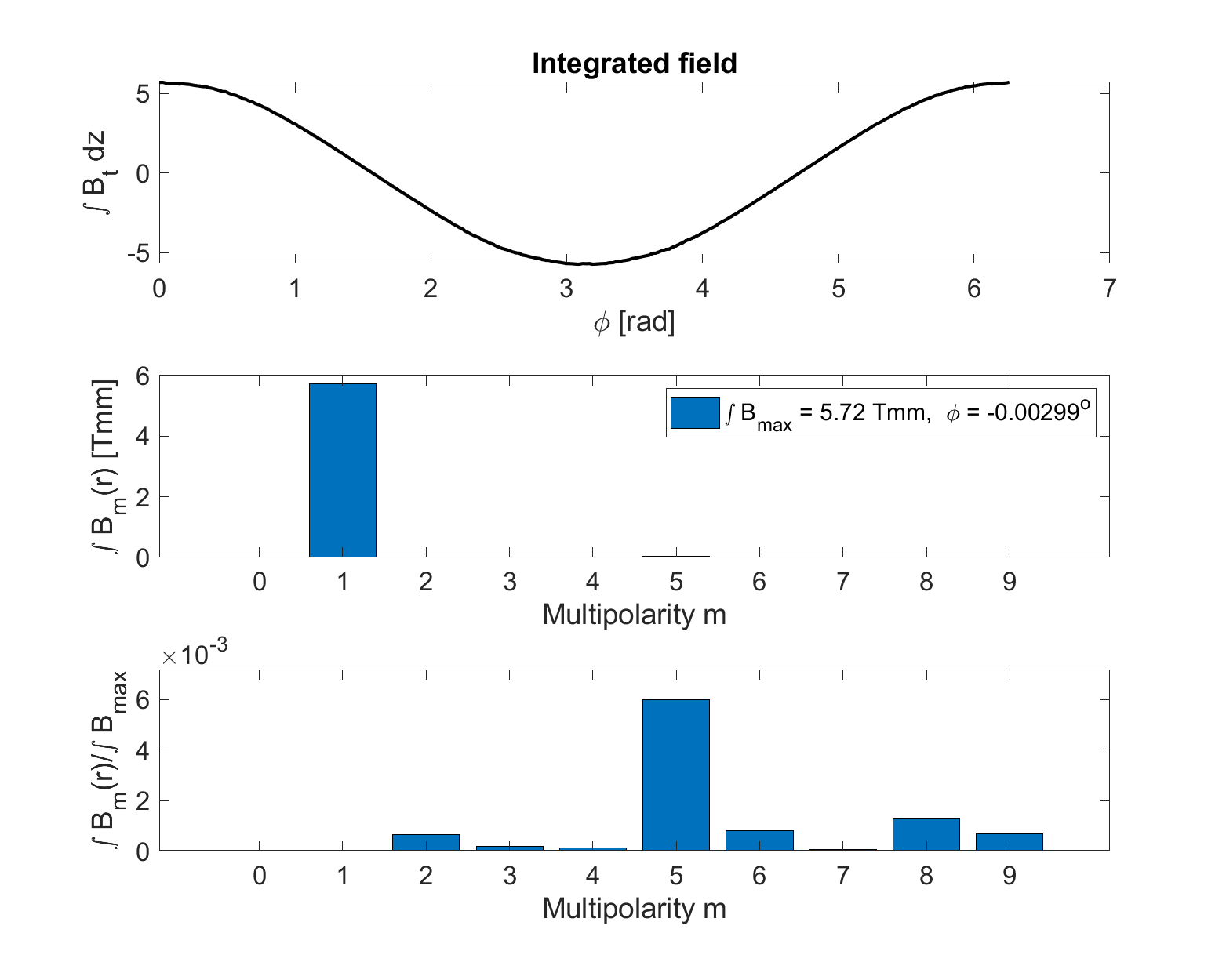}
\includegraphics[width=0.47\textwidth]{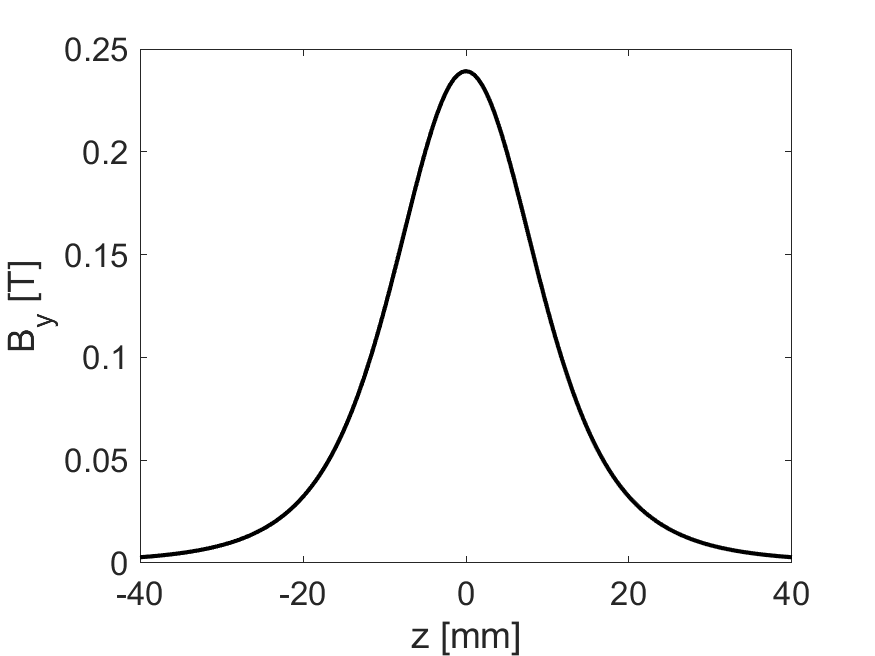}
\end{center}
\caption{\label{fig:dipone}Top left: the eight cubes for a dipole are shown as hollow blue cubes
  and the magnetic field by red arrows. Top right: the vertical field $B_y$ along the
  black dotted line shown on the top-left image. Bottom right: $B_y$ along the
  axis perpendicular through the center of the ring. Bottom left: the tangential
  field on the cyan circle (upper panel), the dominant field harmonic, here for a
  dipole (middle) and the relative strength for the unwanted multipoles (bottom).}
\end{figure}
In~\cite{VZCFE} we calculated the field generated by a rectangular current sheet,
which proved useful to describe permanent-magnet cubes that can be modeled by
four rectangular sheets. In the top-left image in Figure~\ref{fig:dipone} we see
that each of the eight magnets consists of four blue sheets with the green line indicating
the direction of the current, such that we can visualize the cube as a square solenoid
that generates a magnetic field in its inside that is represented by the yellow
arrow, which coincides with the direction of the easy axis of the magnet. In all our
simulations we assume the remanent field of all cubes to be $B_r=1.47\,$T. Inside the
ring-shaped 
assembly with the eight cubes we see red arrows indicating the direction of the field
in the mid-plane. The vertical field $B_y$ along the dotted black line is shown on the
top-right plot, which shows a magnitude of about 0.24\,T with a curved dependence
in the $x$-direction, indicating some higher multipole components. The plot on the
bottom right shows $B_y$ along a line in the $z$-direction. We observe that the
fringe fields roll off slowly and extend far beyond the $\pm 5\,$mm physical size
of the magnets. Finally,
we determine the multipoles of the assembly by integrating the field in the range
$\vert z\vert < 40\,$mm along lines perpendicular to the cyan circle which has a radius
of 5\,mm. We then Fourier-transform the component of the field that is tangential
to this circle and show it on the upper panel on the bottom-left in Figure~\ref{fig:dipone}.
This algorithm, which is explained in~\cite{VZCFE}, gives us the magnitude and angle
of the multipole coefficients of the assembly shown on the middle panel. We see
that the integrated field strength is 5.72\,Tmm  and the lower panel indicates that
the relative magnitude of other multipoles is around $6\times10^{-3}$ with the
decapole contribution ($m=5$) being the largest.
\par
We conclude that the eight-cube dipole magnet reaches a peak field of 0.24\,T, has
an integrated field strength of 5.72\,Tmm and the relative magnitude of unwanted
multipoles on a circle with a 5\,mm radius is below $6\times10^{-3}$.
\section{Quadrupole}
\label{sec:quadone}
%
\begin{figure}[tb]
\begin{center}
\includegraphics[width=0.47\textwidth]{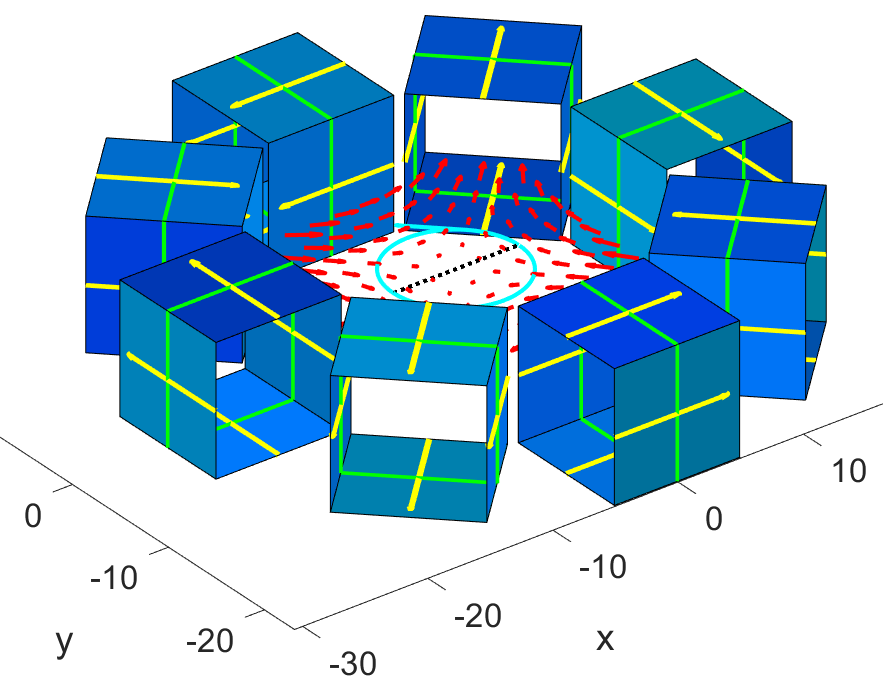}
\includegraphics[width=0.47\textwidth]{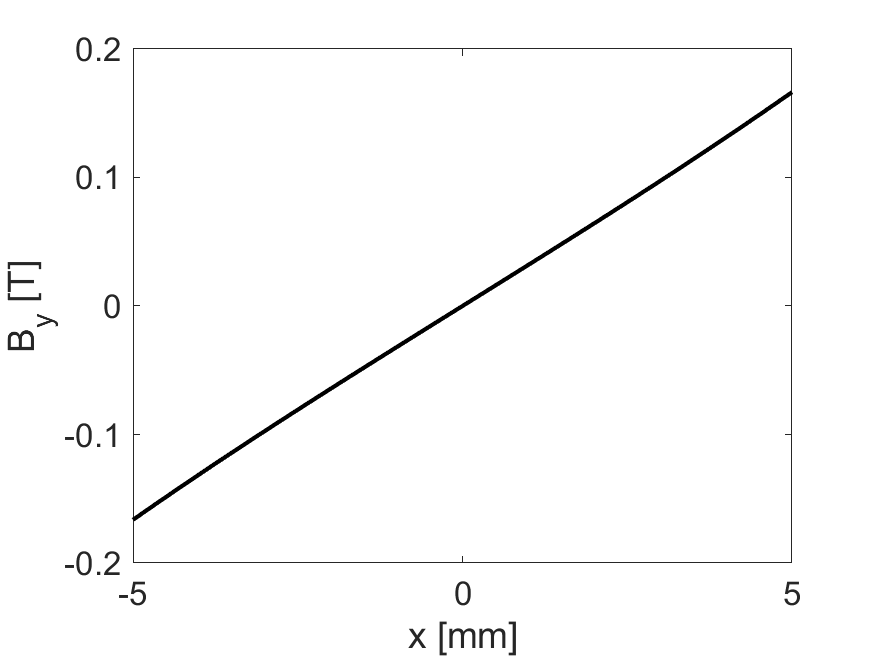}
\includegraphics[width=0.47\textwidth]{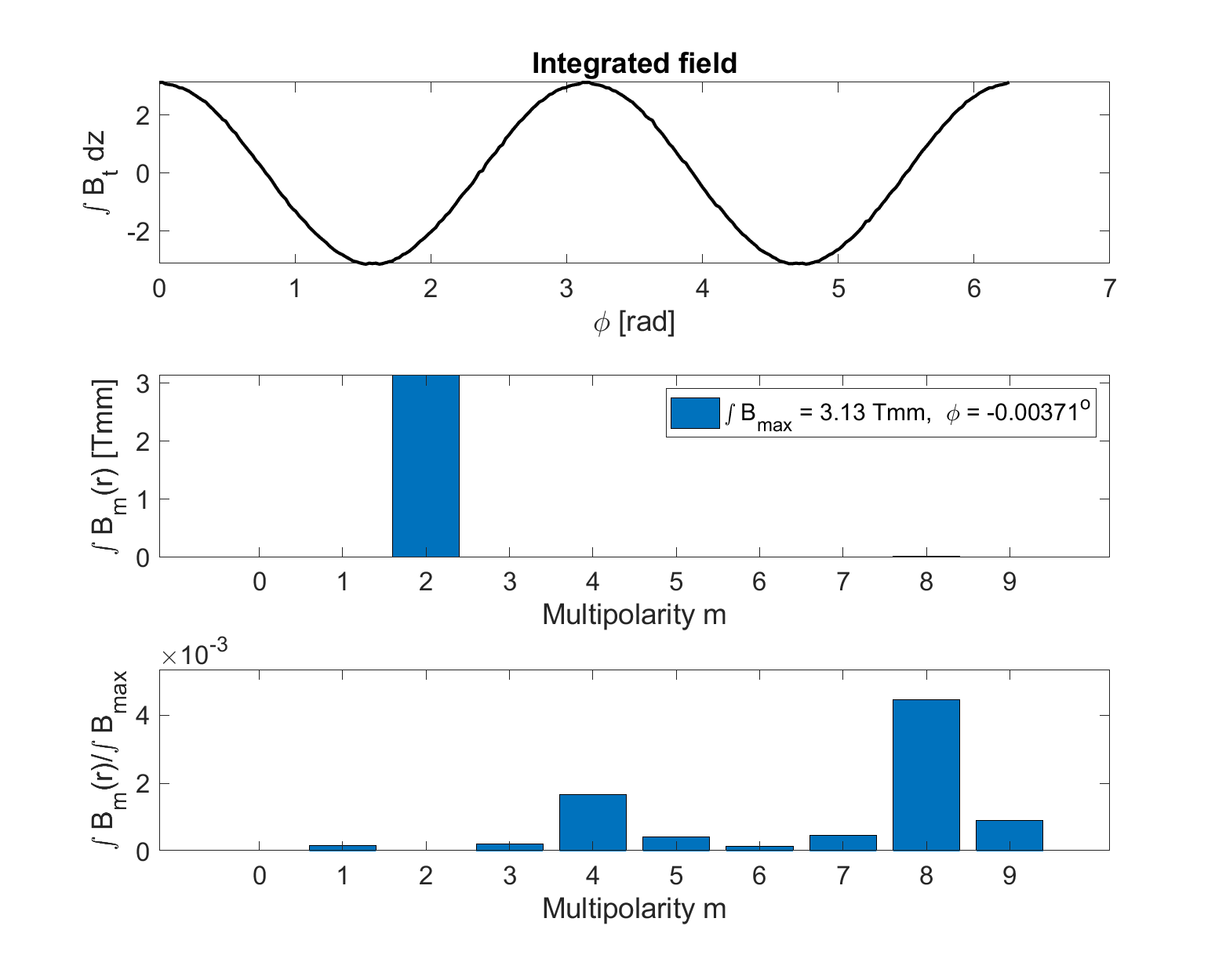}
\includegraphics[width=0.47\textwidth]{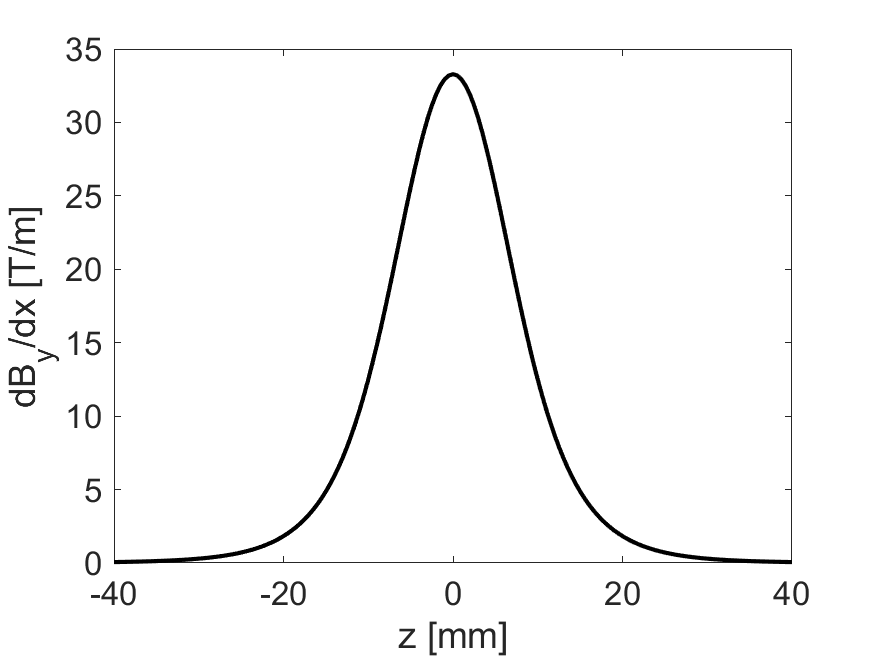}
\end{center}
\caption{\label{fig:quadone}Top left: the eight cubes for a quadrupole are shown as hollow blue cubes
  and the magnetic field by red arrows. Top right: the vertical field $B_y$ along the
  black dotted line shown on the top-left image. Bottom right: the gradient
  $dB_y/dx$ along the axis perpendicular through the center of the ring. Bottom left:
  the tangential field on the cyan circle (upper panel), the dominant field harmonic,
  here for a quadrupole (middle) and the relative strength for the unwanted multipoles
  (bottom).}
\end{figure}
The field of the quadrupole assembly corresponding to frame from the right in
Figure~\ref{fig:frames} is shown in Figure~\ref{fig:quadone}. The top-left
image shows the eight cubes with the red arrows indicating the field in the
mid-plane, the black dotted line and the cyan circle on which the multipoles
are calculated. The top-right plot shows $B_y$ along the black dotted line
and indicates a linearly rising field along the $x$-axis that is
characteristic for a quadrupole. From the slope we determine a gradient in the
mid-plane of $32.7\,$T/m. The gradient along the $z$-axis in the middle of the
assembly is visible on the bottom-right plot. We observe that even the fringe
fields of the quadrupole extend beyond its physical size of $\pm 5\,$mm, but
decays faster than those of a dipole. The image with the three panels on the
bottom-left shows the tangential field component around the azimuth, which clearly
displays two oscillations, indicating  a quadrupolar field. This is also verified
on the middle panel, which shows an integrated field 3.13\,Tmm, which, when dividing
by the 5\,mm radius of the circle, results in an integrated gradient of 0.63\,T.
The lower panel shows that all other multipoles contribute less than $4\times 10^{-3}$
in magnitude.
\section{Solenoids}
\label{sec:solone}
%
\begin{figure}[tb]
\begin{center}
\includegraphics[width=0.47\textwidth]{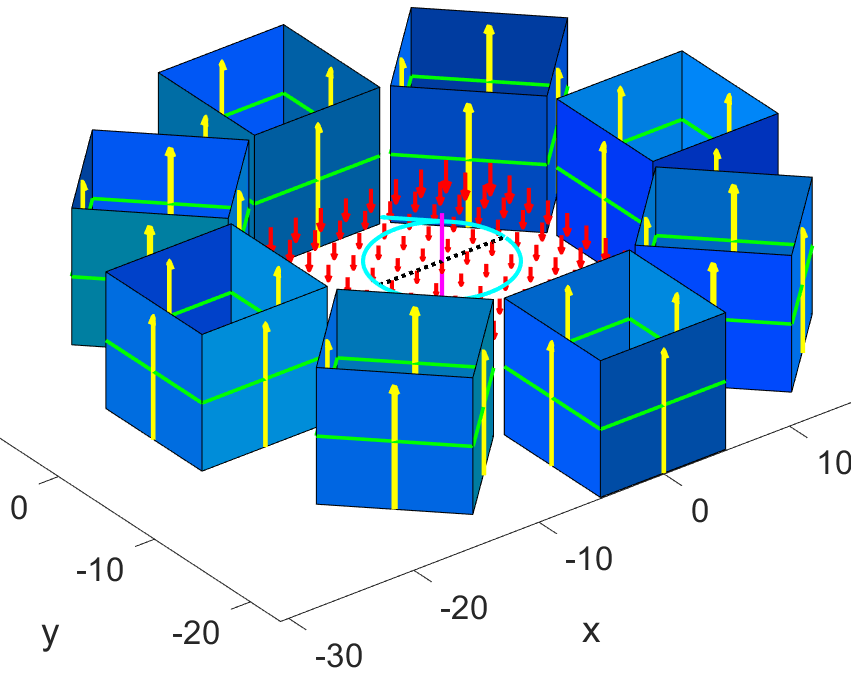}
\includegraphics[width=0.47\textwidth]{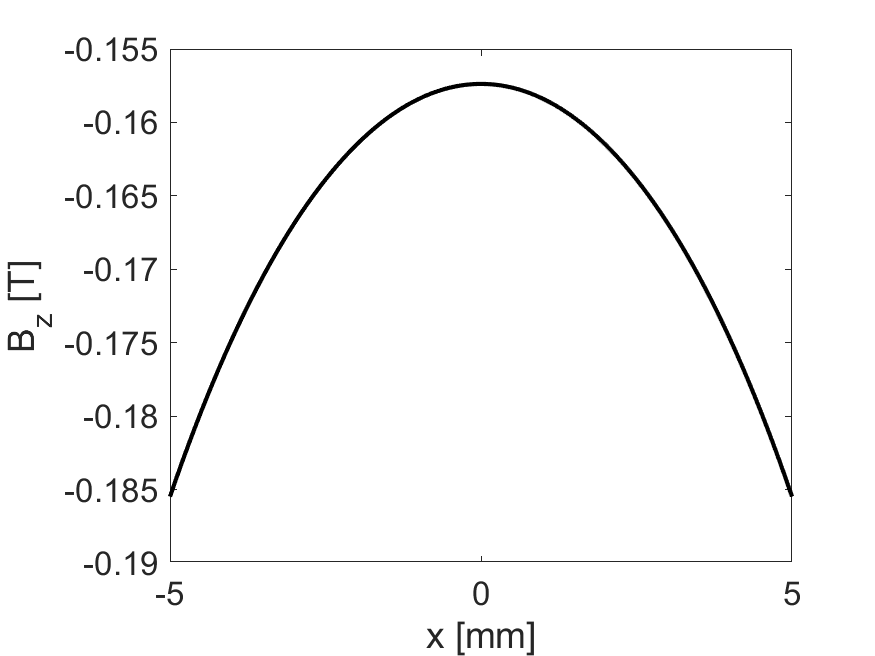}
\includegraphics[width=0.47\textwidth]{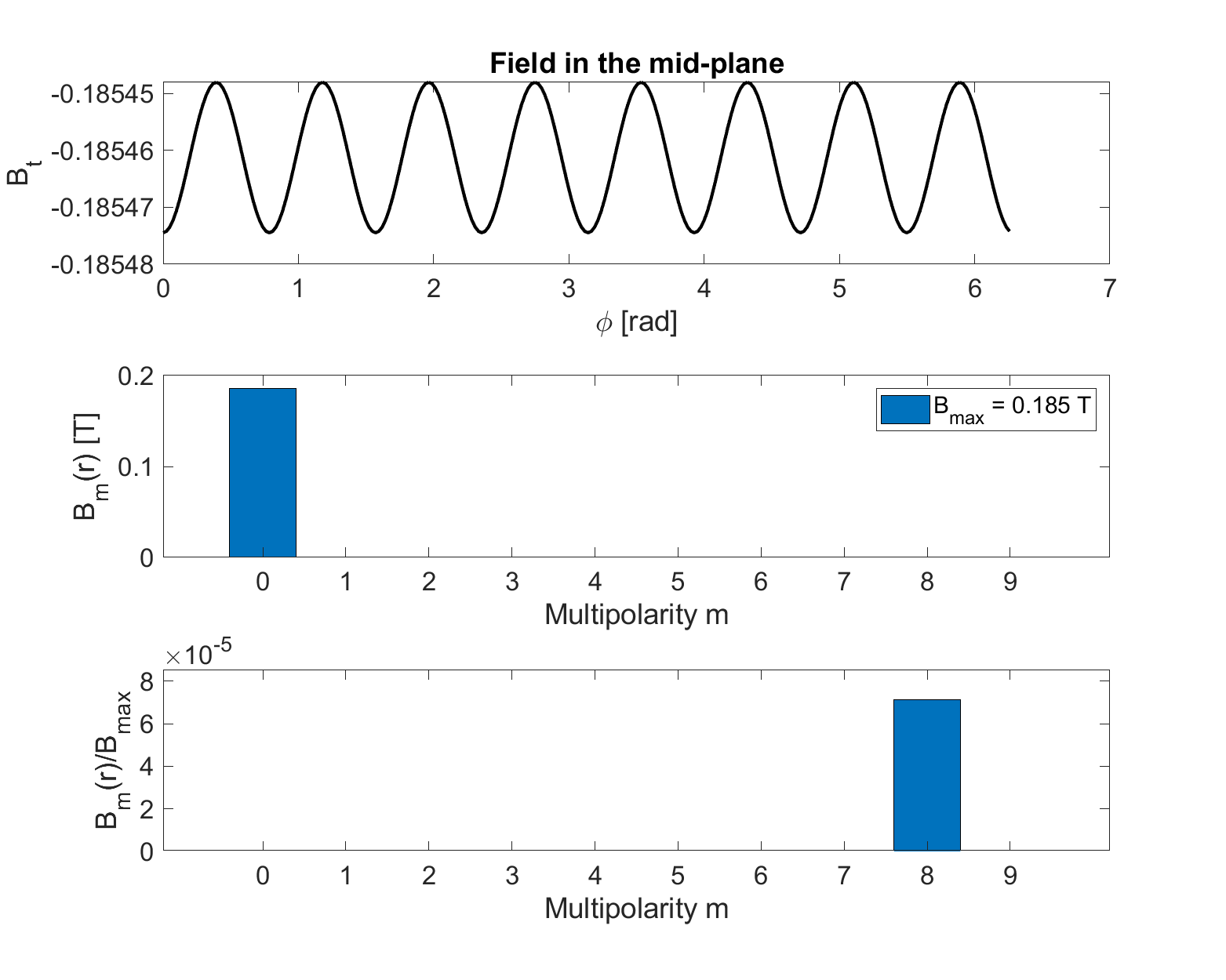}
\includegraphics[width=0.47\textwidth]{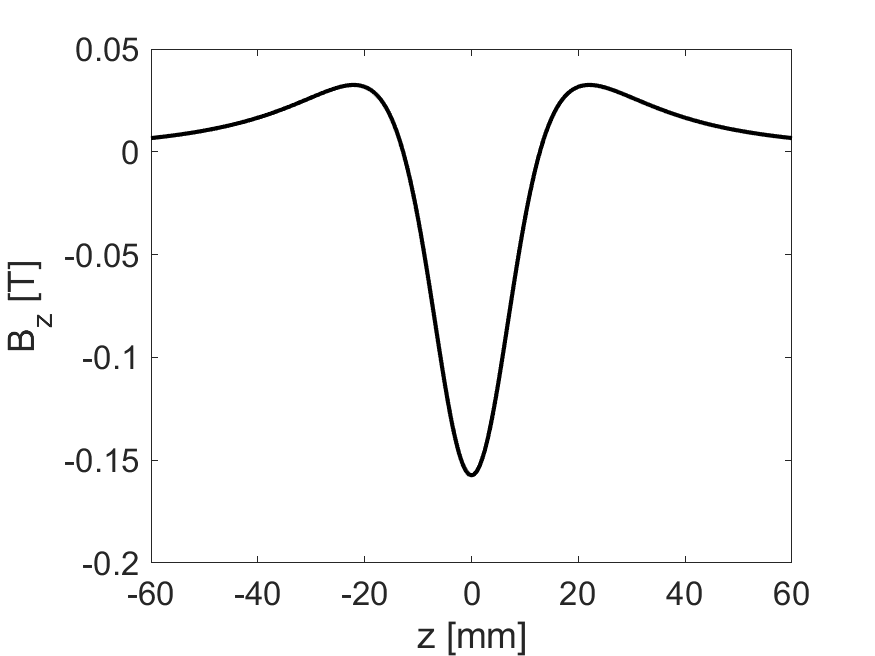}
\end{center}
\caption{\label{fig:solone}Top left: the eight cubes for an axial solenoid are shown as
  hollow blue cubes and the magnetic field by red arrows. Top right: the axial field $B_z$
  along the black dotted line shown on the top-left image. Bottom right: $B_z$ along the
  axis perpendicular through the center of the ring. Bottom left: the axial field $B_z$
  on the cyan circle (upper panel), the dominant field harmonic, here constant (middle)
  and the relative strength for the unwanted harmonics.}
\end{figure}
It turns out that the frame for the quadrupoles can also be used to construct both
axial and radial solenoids; for an axial solenoid
we only have to insert all cubes with their easy axis pointing along the $z$-axis. The
top-left image in Figure~\ref{fig:solone} illustrates this. Note also the black dotted
line and the cyan circle in the midplane as well as the solid magenta line to illustrate
the $z$-axis. The top-right plot shows
the $z$-component $B_z$ of the magnetic field in the midplane along the black dotted
line visible on the top-left image. We see that is has a parabolic shape---the
magnitude of the field closer to the permanent magnets is higher.
We see that $B_z$ points opposite the direction of the easy axis of the cubes
and reaches about $-0.16\,$T in the center of the assembly. On the lower-right plot
we also observe
that $B_z$, after being negative in the center of the assembly, becomes positive
with a magnitude of 0.03\,T around $z=\pm20\,$mm.
On the three-panel image on the
bottom-left we explore the impact of approximating the solenoid by eight cubes.
The upper panel shows $B_z$ in the midplane of the assembly around the cyan circle
which has a radius of $5\,$mm. We observe eight oscillations as we move around the
azimuth $\phi$, whose amplitude, however is rather small, less than $10^{-4}$ time
the magnitude of the average $B_z$, which is $0.185\,T$ at $r=5\,$mm. We conclude
that, even though the field quality is limited, constructing solenoids is feasible
within our framework.
\par
\begin{figure}[tb]
\begin{center}
\includegraphics[angle=270,origin=c,width=0.44\textwidth]{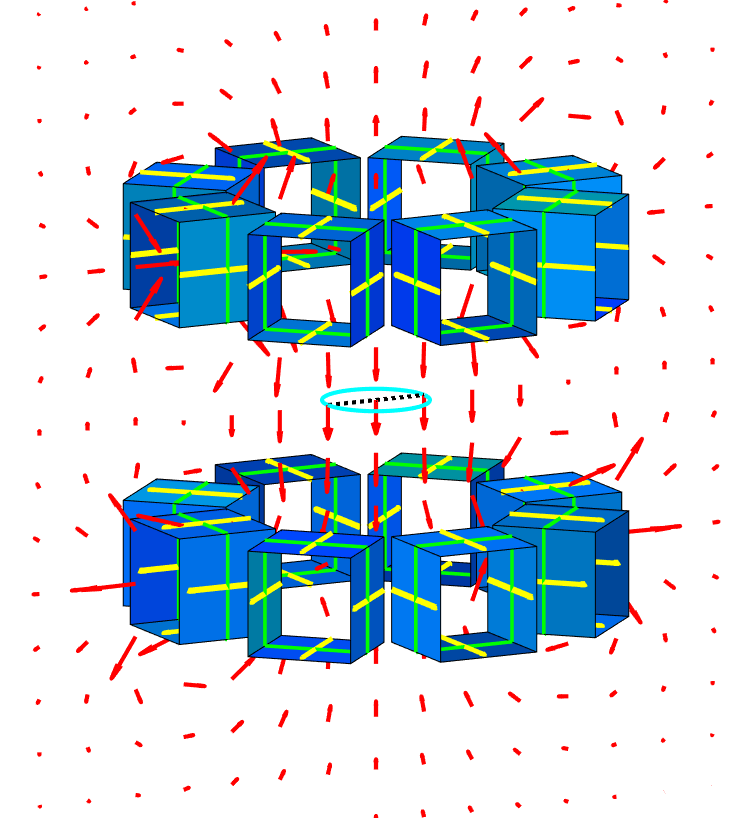}
\includegraphics[width=0.55\textwidth]{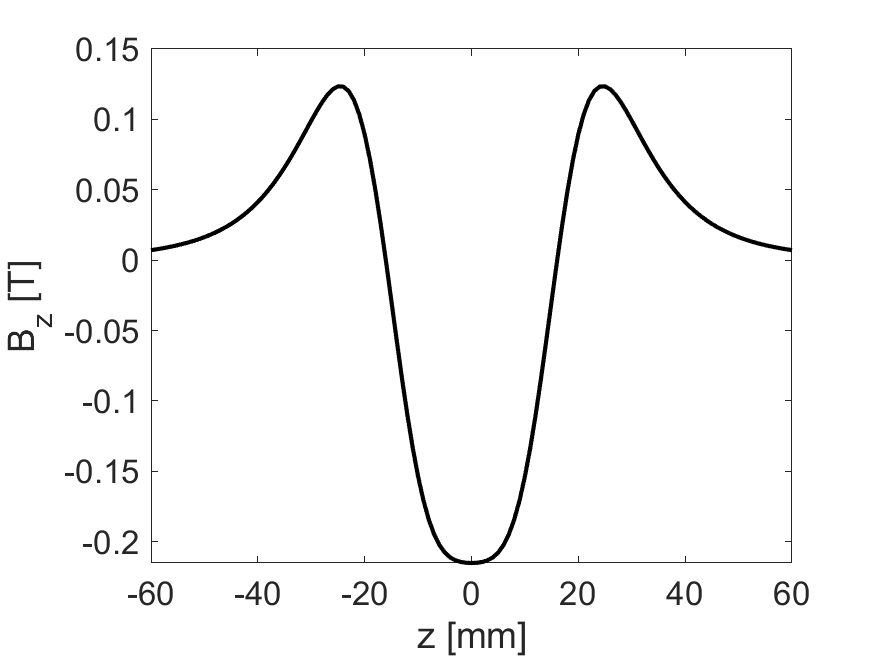}
\end{center}
\caption{\label{fig:solrad}Left: A radial solenoid made of two rings and the field illustrated by
  the red arrows. Right: the axial field $B_z$ along a line through the center of the two rings.}
\end{figure}
Likewise we can construct a radial solenoid by rotating the cubes such that their
easy axis either points inwards or outwards and construct one solenoid from two
rings, as shown in Figure~\ref{fig:solrad}. The easy axes of the left eight-cube
rings point inward and those of the right ring point outward. The magnetic
field is illustrated by the red arrows and points from the right towards the left
assembly along the negative $z$-axis in the intermediate region. The plot on the
right-hand side in Figure~\ref{fig:solrad} shows $B_z$ along the $z$-axis, where
$z=0$ corresponds half way between the two assemblies whose position is indicated
by the blue dashed boxes. We see that the maximum value exceeds 0.2\,T in the center
and and reaches about 0.12\,T just outside the two assemblies. The homogeneity of
the field is comparable to that of the axial solenoid close to the cubes, and
better in the intermediate region.
\section{Adjustable dipole}
\label{sec:vardip}
%
\begin{figure}[p]
\begin{center}
\includegraphics[width=0.47\textwidth]{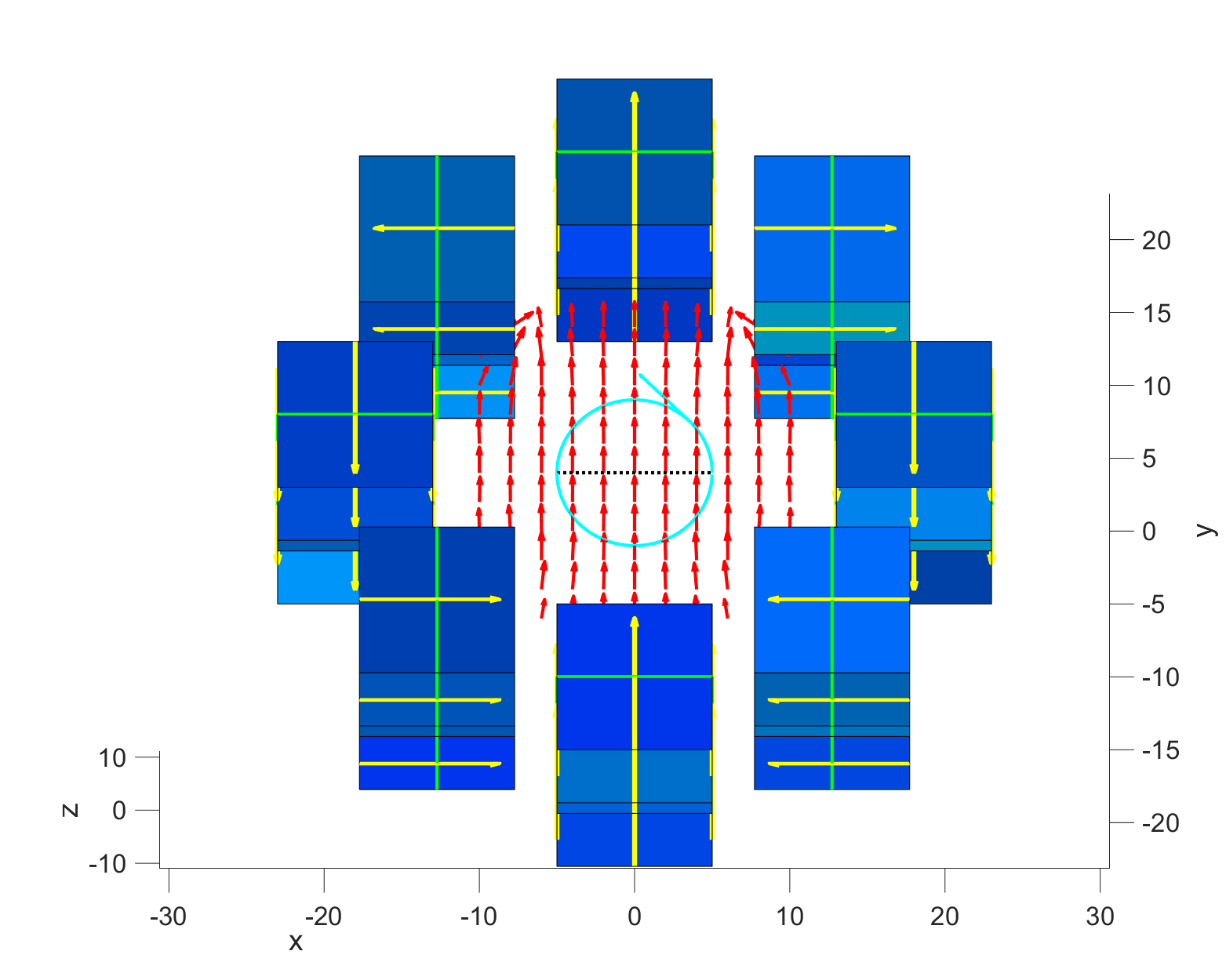}
\includegraphics[width=0.47\textwidth]{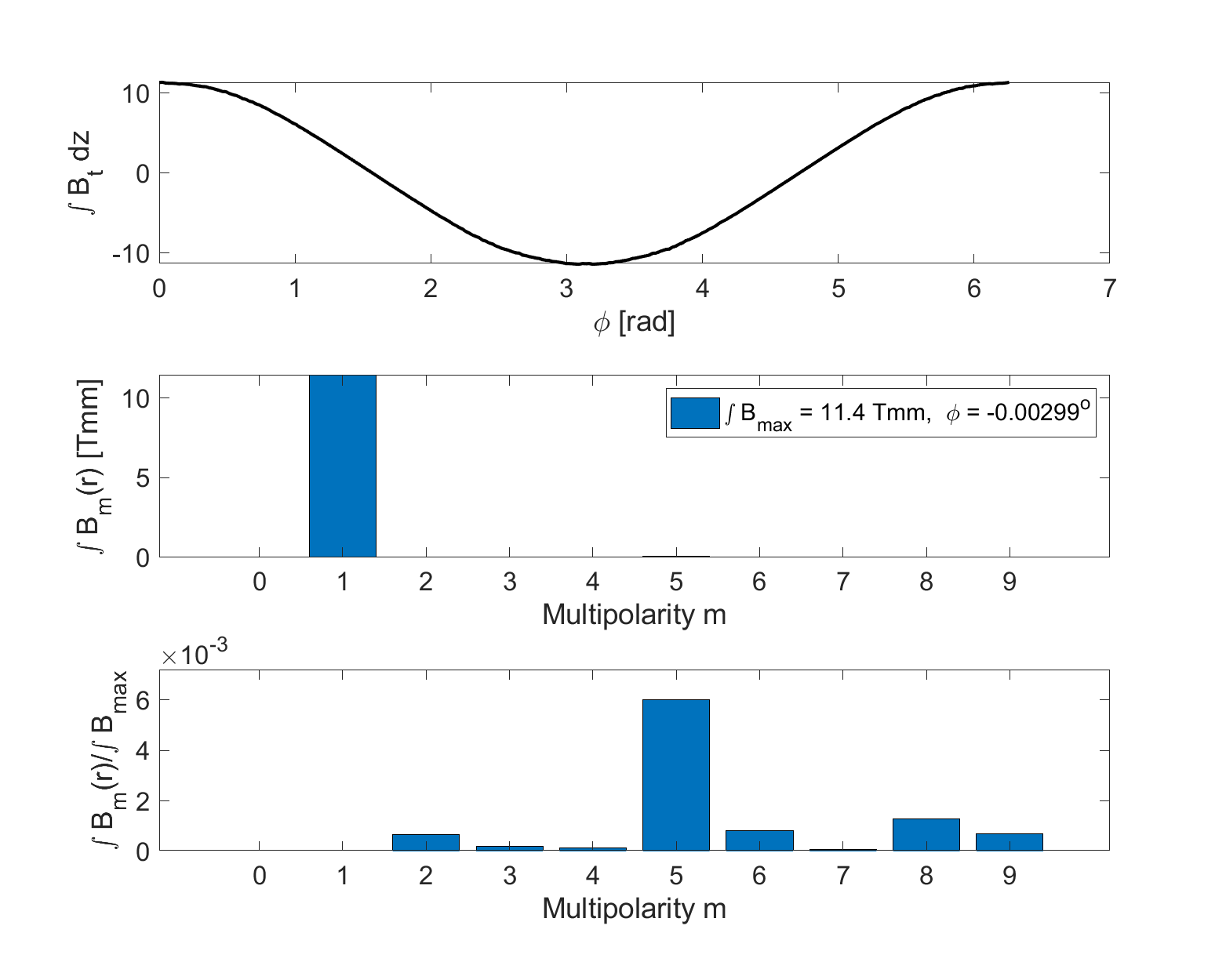}
\includegraphics[width=0.47\textwidth]{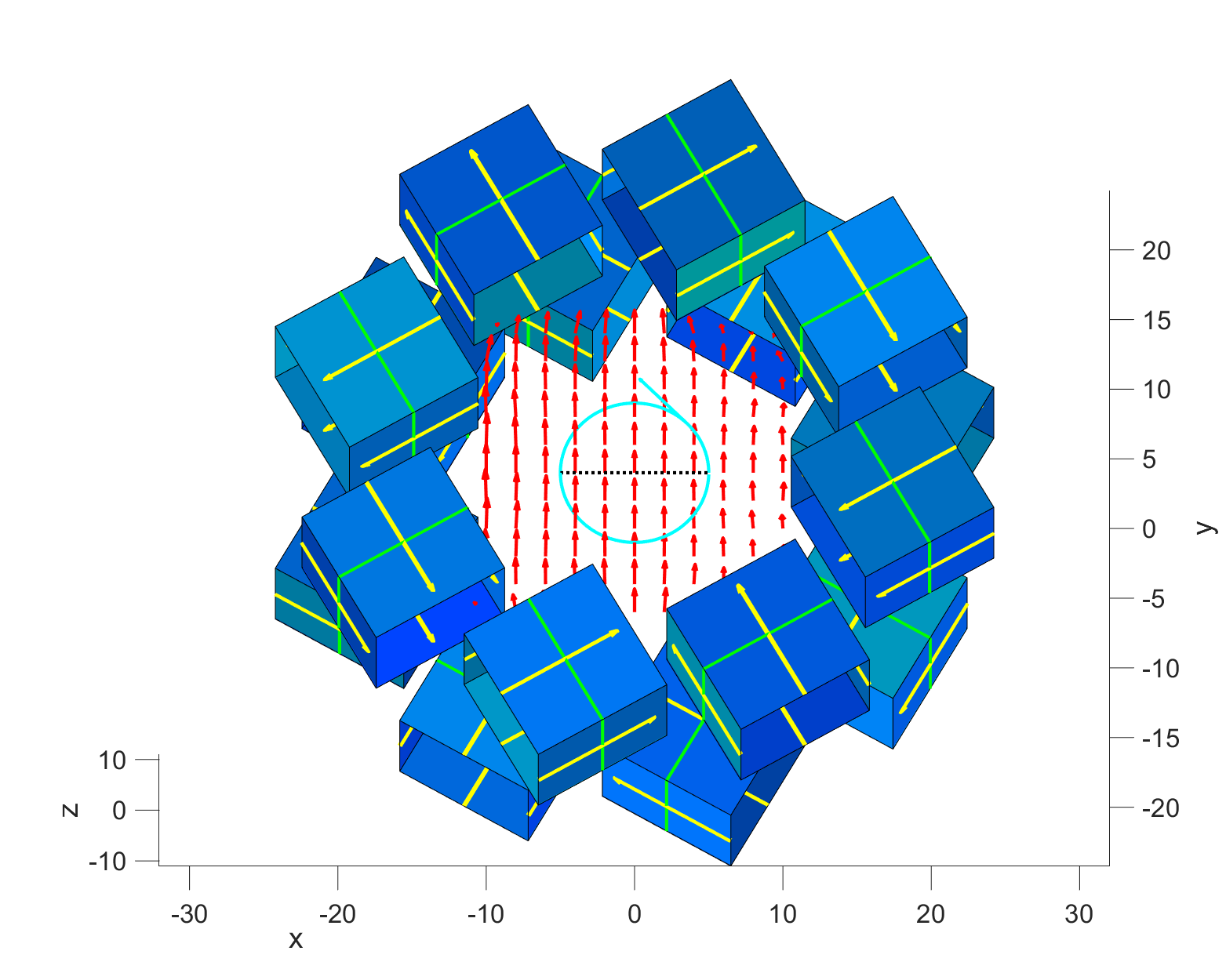}
\includegraphics[width=0.47\textwidth]{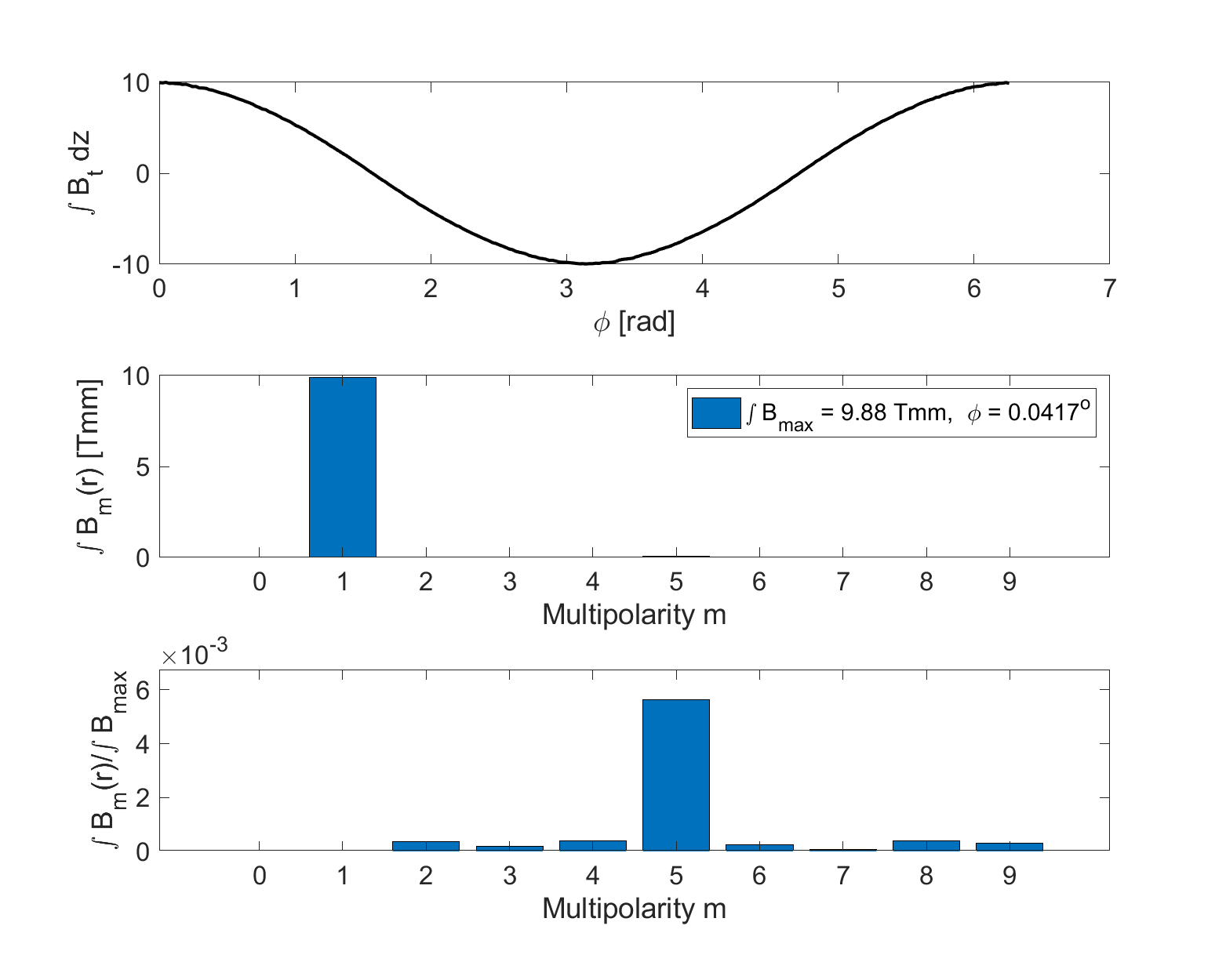}
\includegraphics[width=0.47\textwidth]{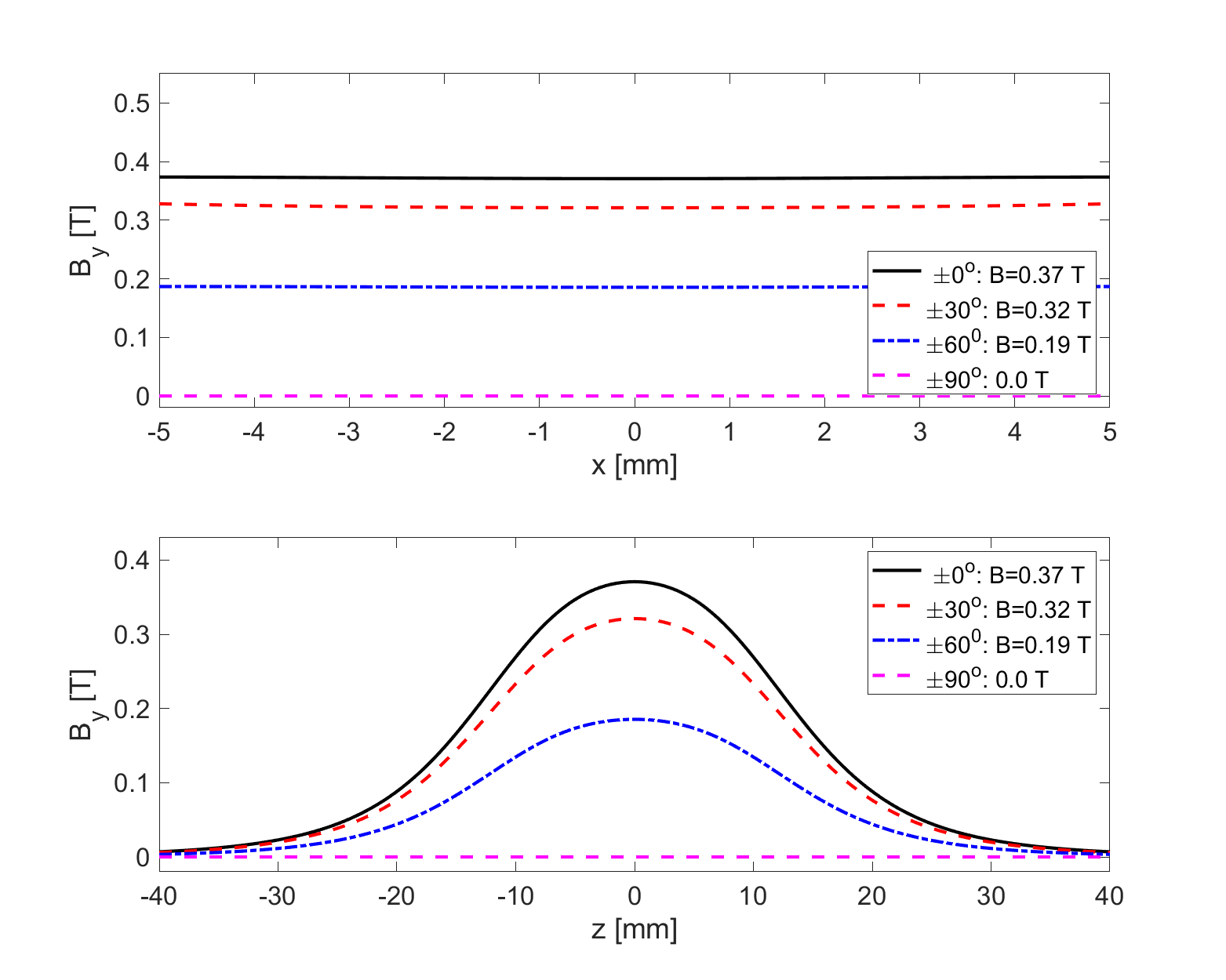}
\includegraphics[width=0.47\textwidth]{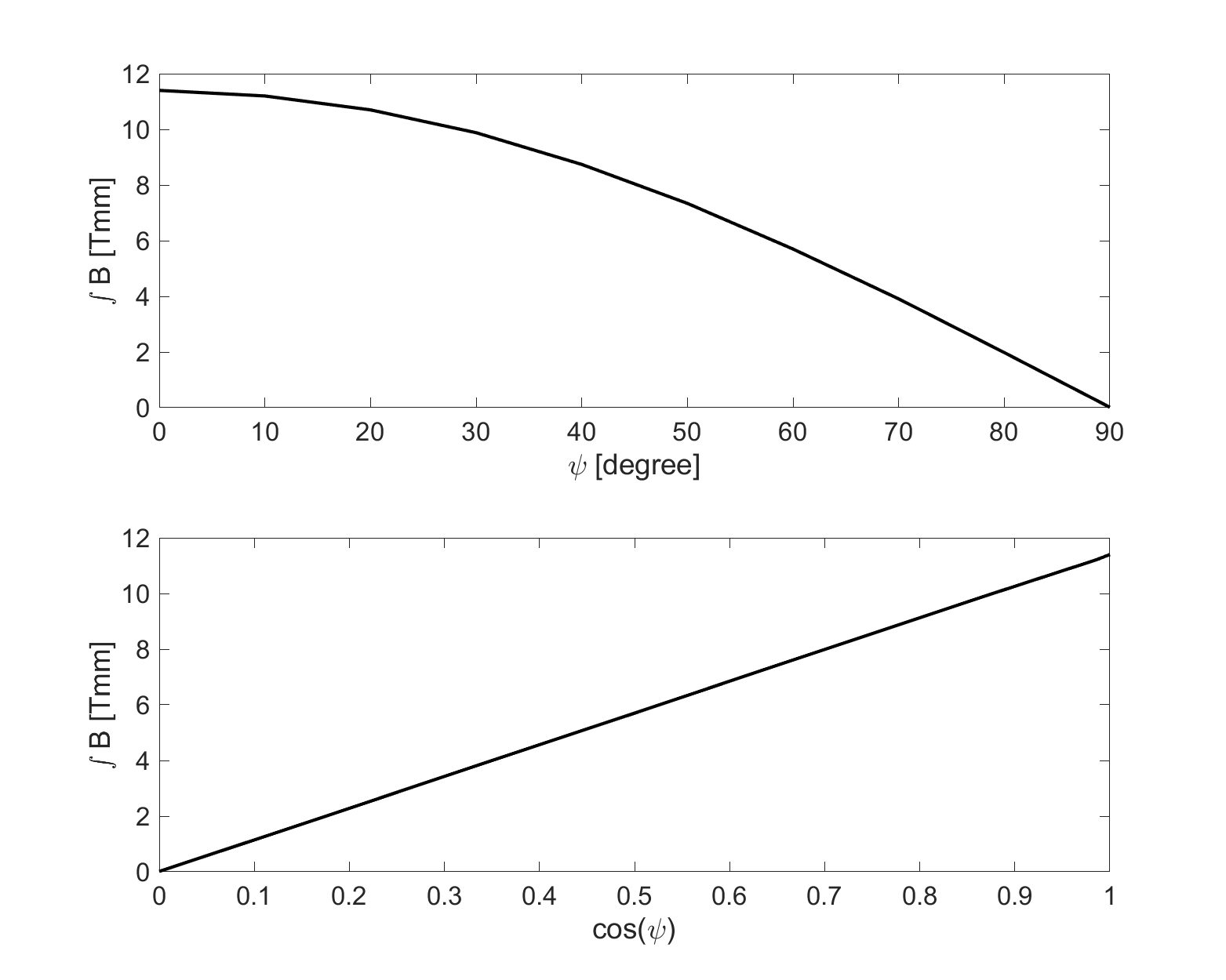}
\end{center}
\caption{\label{fig:dip}Top left: Two rings for a variable dipole aligned on top of each
  other. Top right: the
  tangential field on the cyan circle (upper), the dominant field harmonic (middle), and
  the relative strength of the unwanted harmonics (bottom). Middle row: the corresponding
  images for rings rotated with $\psi=30^o$. The integrated strength is reduced, but the
  unwanted harmonics are similar to the plot above. Bottom left: the fields along the black
  dotted line for different rotation angles $\psi$ (upper) and $B_y$ along a line through
  the center of the two rings. Bottom right: the integrated field plotted versus $\psi$
  (upper) and $\cos\psi$ (lower).}
\end{figure}
Inspired by the observation that the fields of two very short magnets with opposite
polarity almost cancel and the fact that the fringe fields of the dipole from
Section~\ref{sec:dipone} extend far beyond the physical length of the magnet, we
place two dipole rings very close to each other (2\,mm between the cubes) and then rotate one
with respect to the other. The images on the left-hand side in the top and middle
row of Figure~\ref{fig:dip} illustrate the geometry of these assemblies. On the
top left the two eight-cube rings are aligned and on the one below the upper
ring is rotated by $\psi=30^o$ counter-clockwise while the lower ring is rotated
by $30^o$ clockwise.
\par
The three-panel images towards the right of the respective geometries illustrate the
integrated multipole content of the assemblies. In the upper panel on the top-right
image we see that the tangential field component of the field integral shows
a sinusoidal pattern that is responsible for dominant dipole component visible on the
middle panel. The integrated strength of $11.4\,$Tmm of this assembly is twice
that of the single magnet from Figure~\ref{fig:dipone}, because the fields superimpose
linearly in ideal permanent-magnet assemblies. Furthermore, the lower panel shows that
the dominant higher multipole is the decapole ($m=5$) component whose contribution is
$6\times 10^{-3}$ smaller than the main component. The right-hand image in the middle
row shows the integrated field of the assembly with the two rotated magnets. The
tangential field shown on the upper panel is also sinusoidal, albeit with a reduced
amplitude of $9.88\,$Tmm, which is also given in the legend of the middle panel with
the Fourier-harmonics. Even in the rotated geometry the dipole contribution is dominant
and the relative contributions of the unwanted harmonics, shown on the lower panel are
below $6\times 10^{-3}$.
\par
That the dominant effect of rotating two eight-cube magnets with opposite angles
$\psi$ only changes the integrated field from $11.4$ to $9.88\,$Tmm but does not
adversely affect the unwanted multipoles encourages us to explore the dependence
of the field integral for different rotation angles $\psi$. The image on the
left-hand side in the bottom row shows the vertical field $B_y$ along the $z$-axis
on the lower panel for $\psi=0^o, 30^o, 60^o,$ and $90^o$. We see that rotating
the two eight-cube rings reduces the peak field in the center of the magnet
from $0.37\,$T down to zero. The upper panel shows $B_y$ on the $x$-axis along
the black dotted line and we see that the field is rather constant, only the
amplitude is different for different values of~$\psi$. The lower panel shows
the fringe fields of $B_y$ along $z$ for the four values of~$\psi$.
\par
Finally, we systematically vary $\psi$ in steps of $10^o$ and show the integrated
field component (or the amplitude of the Fourier harmonic for $m=1$) as a function
of $\psi$ on the upper panel in the image on the right-hand side in the bottom
row of Figure~\ref{fig:dip}. We observe that the integrated field decreases in
a smooth  curve from 11.4\,Tmm down to zero at $\psi=90^o$, where the easy axes
of cubes on top of each other have opposite polarities and cancel one another.
On the lower panel we show the same data, but plot versus $\cos\psi$ instead.
We find that now the integrated field strength shows a linear dependence on
$\cos\psi$, which indicates that the integrated field is given by twice the
integrated field of one eight-cube ring times $\cos\psi$. But this is just
the projection of the maximum field of the rings onto the vertical axis. The
field components along the vertical axis add, but along the horizontal axis
they cancel. This will make tuning and setting up of such double-dipole
assemblies rather straightforward.
\section{Adjustable quadrupole}
\label{sec:varquad}
%
\begin{figure}[p]
\begin{center}
\includegraphics[width=0.49\textwidth]{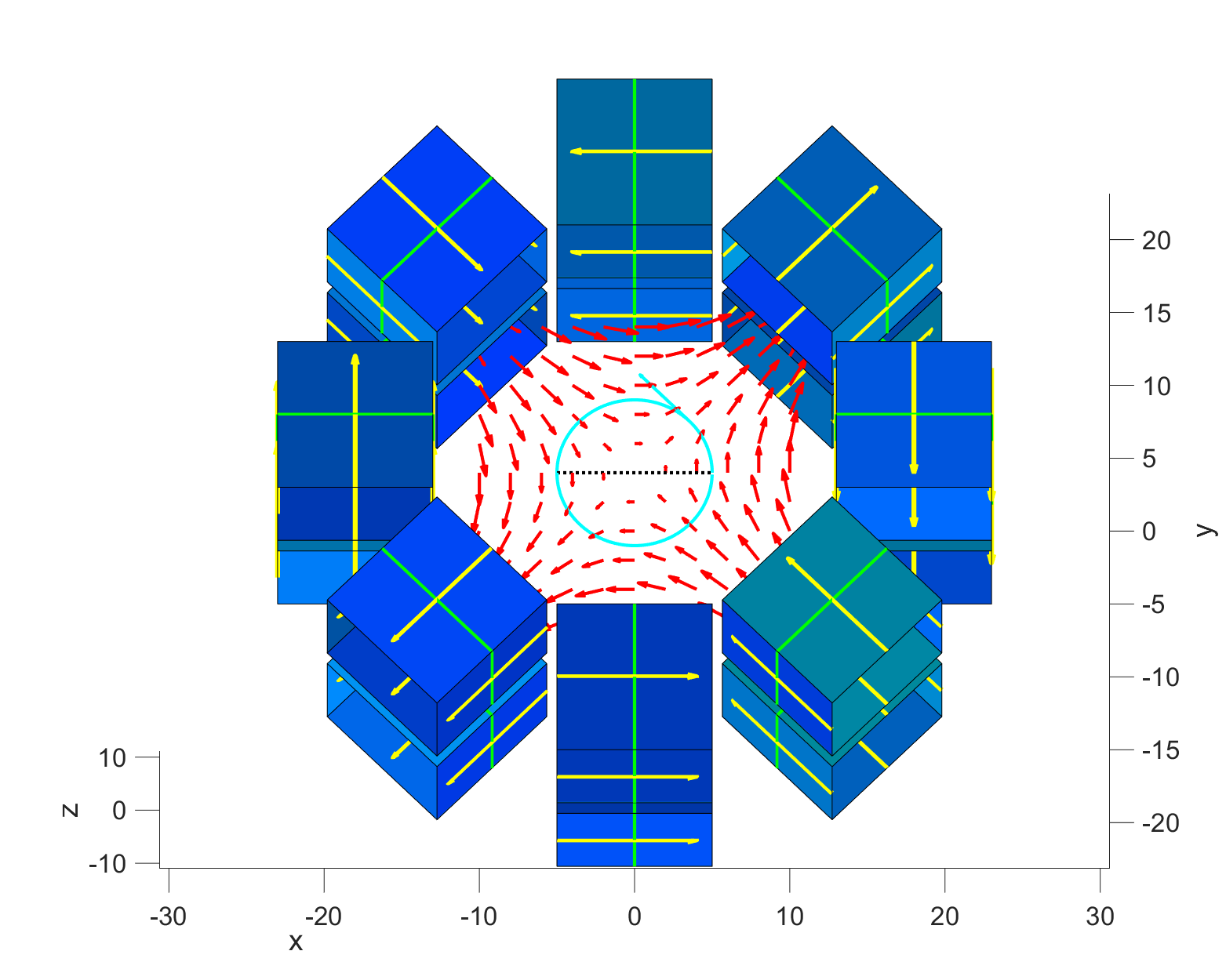}
\includegraphics[width=0.49\textwidth]{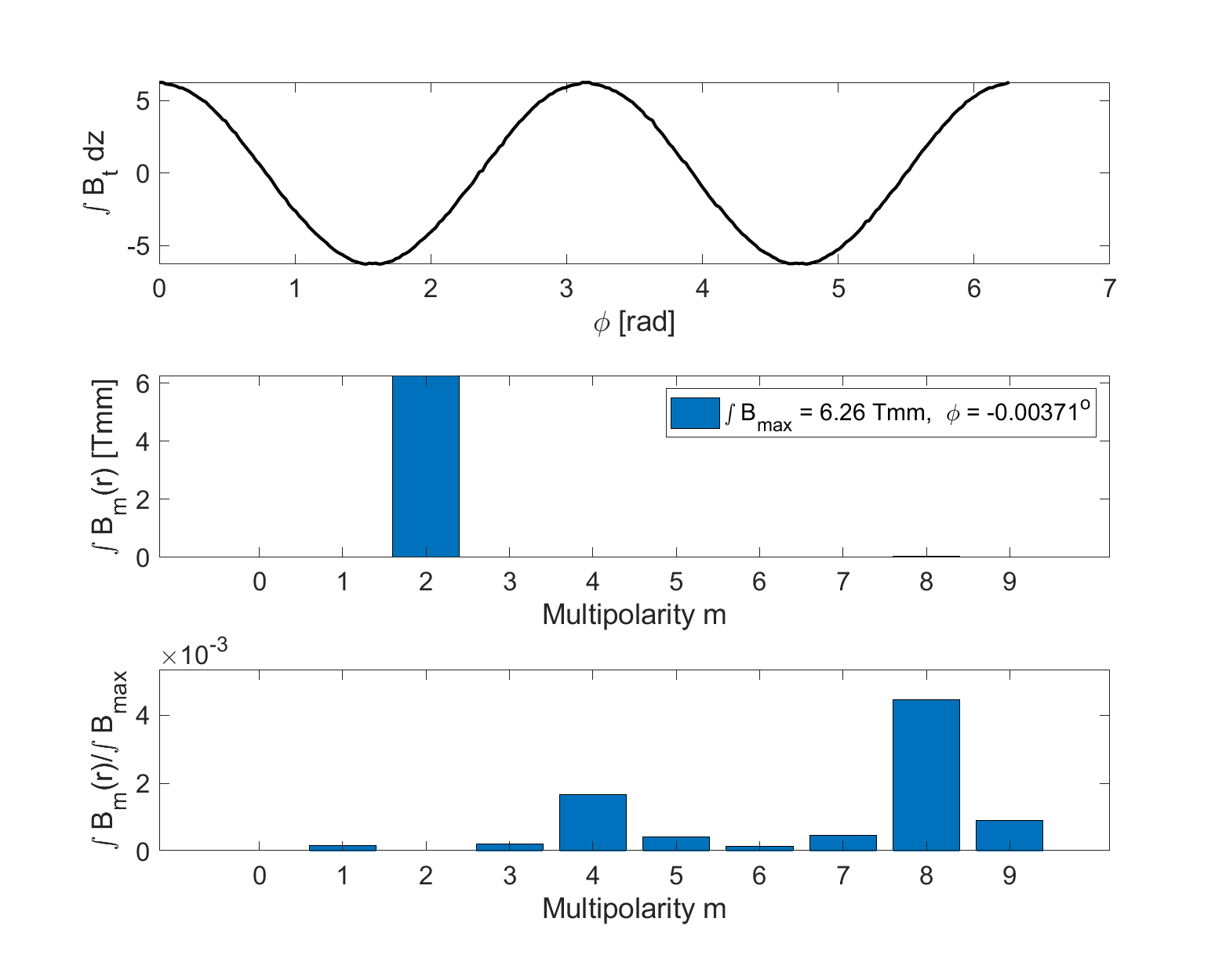}
\includegraphics[width=0.49\textwidth]{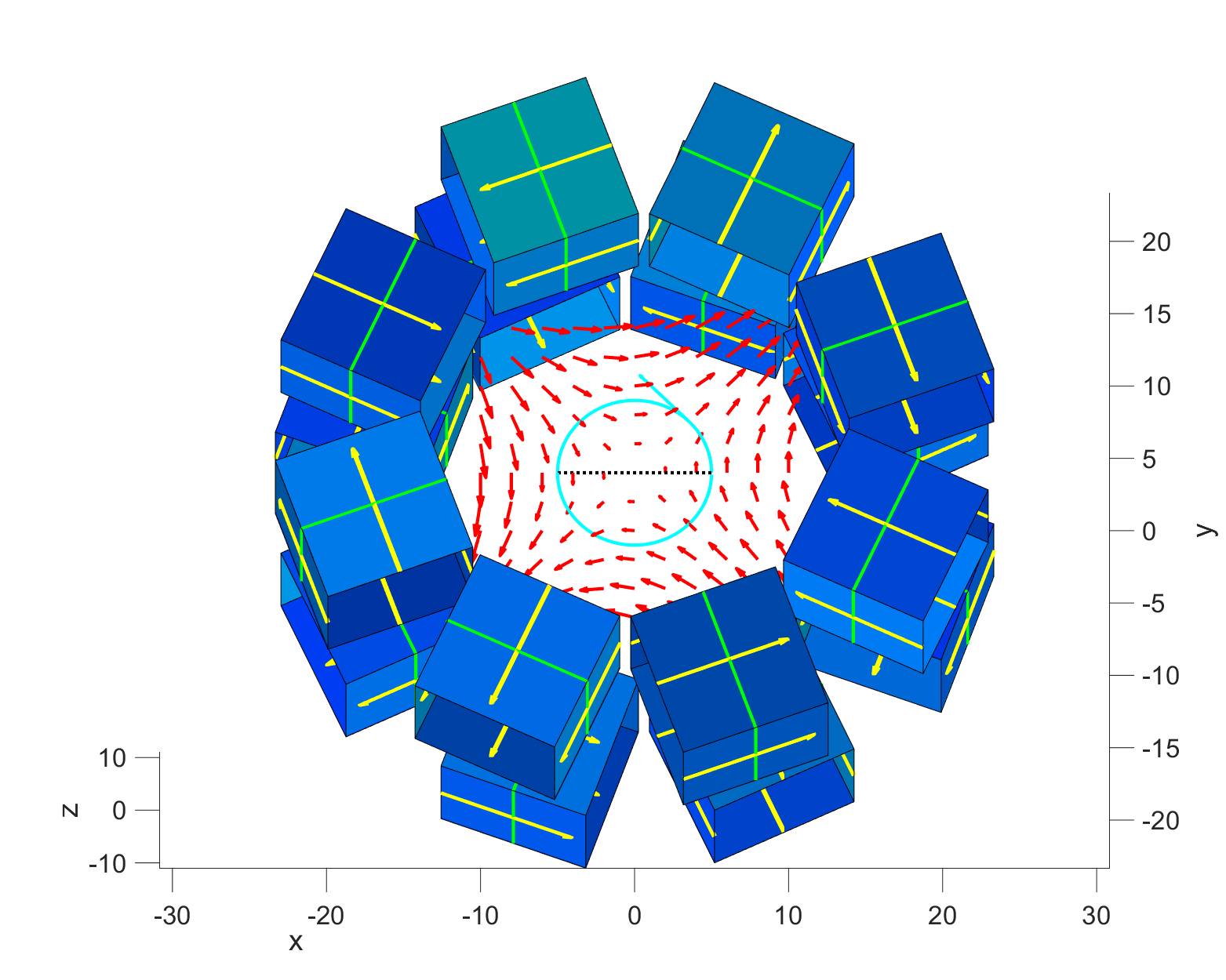}
\includegraphics[width=0.49\textwidth]{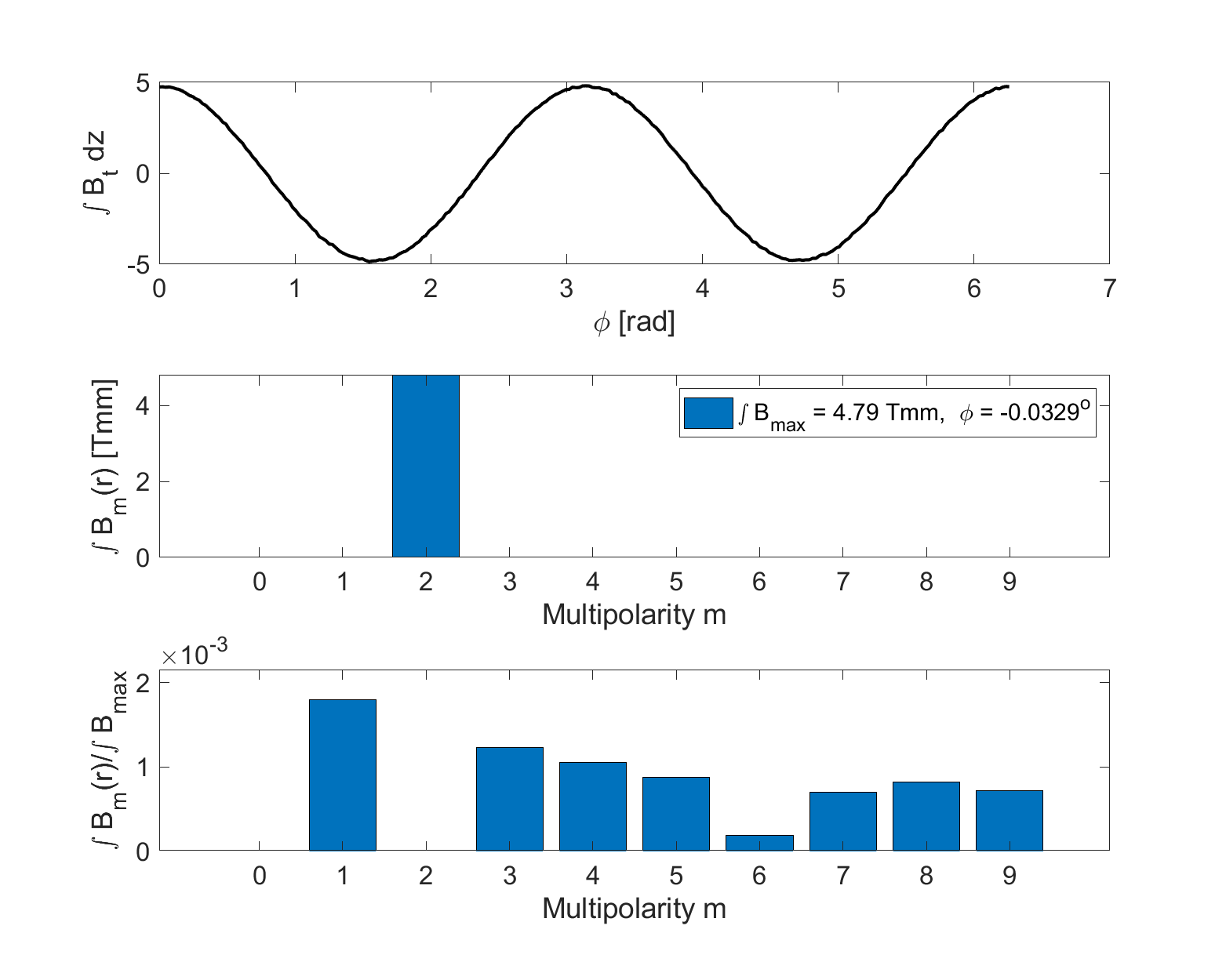}
\includegraphics[width=0.49\textwidth]{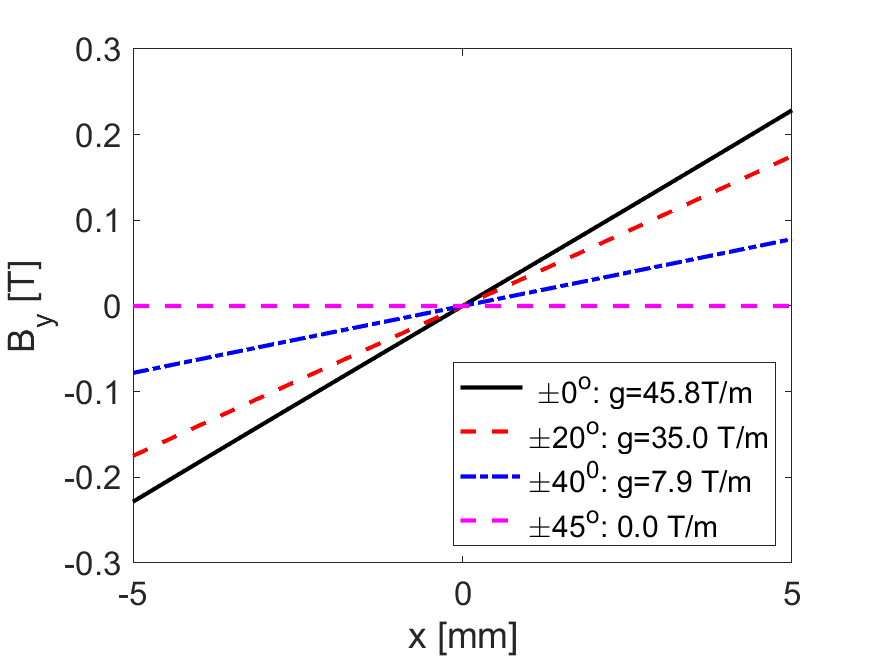}
\includegraphics[width=0.49\textwidth]{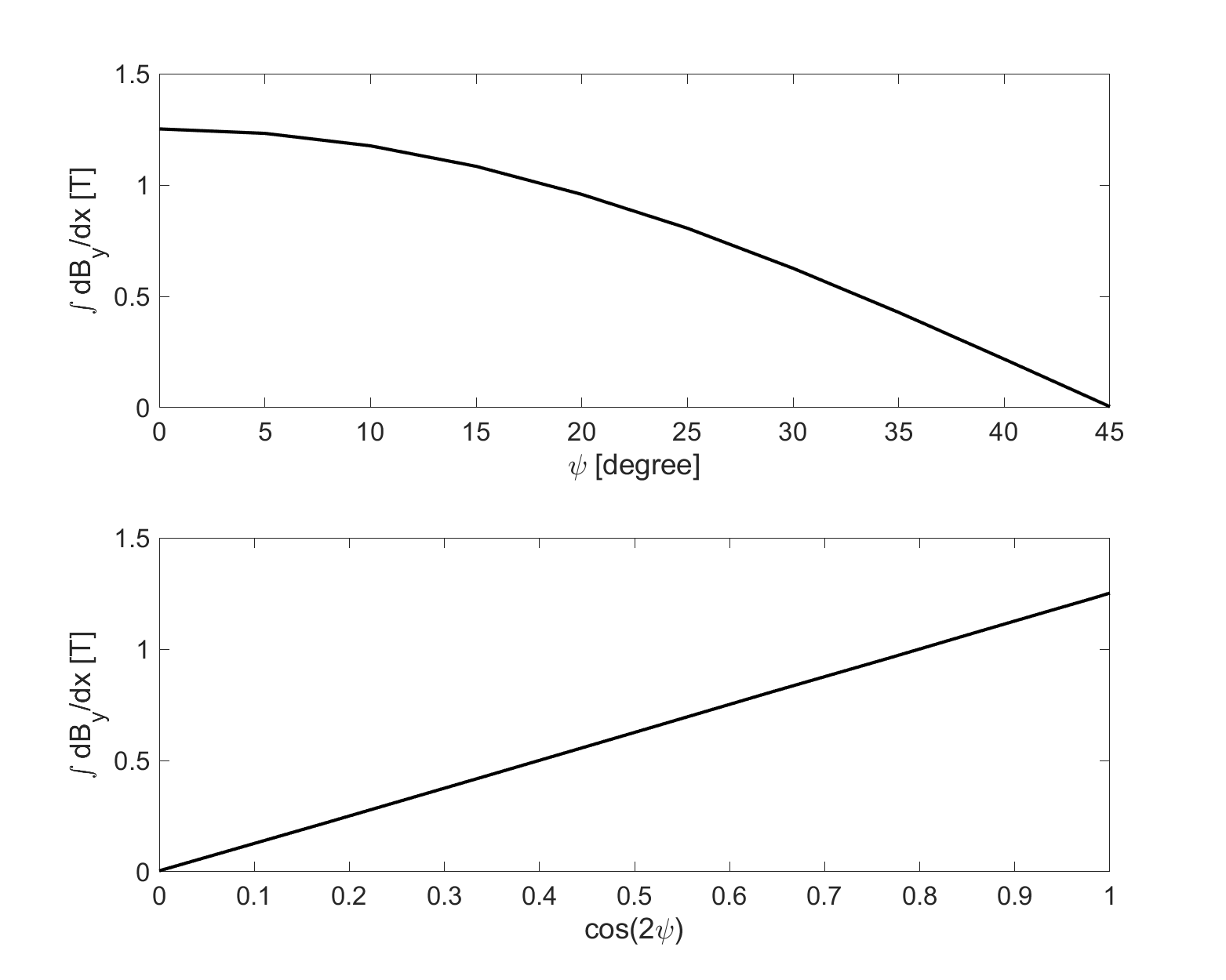}
\end{center}
\caption{\label{fig:quad}Top left: two rings for a variable quadrupole aligned on top
  of each other. Top right: the tangential field on the cyan circle (upper), the dominant
  field harmonic (middle), and the relative strength of the unwanted harmonics (bottom).
  Middle row: the corresponding images for rings rotated with $\psi=20^o$. The amplitude
  is reduced, but the unwanted harmonics are similar to the plot above. Bottom left: the
  gradient along the black dotted line for different rotation angles $\psi$. Bottom right:
  The integrated gradient as a function of the rotation angle $\psi$ (upper) and as a
  function of $\cos2\psi$.}
\end{figure}
It is certainly no surprise that the same idea also works with quadrupoles. On the
top-left image is Figure~\ref{fig:quad} we show the geometry of two aligned ($\psi=0$)
eight-cube quadrupoles on top of each other with 2\,mm space between the rings. On the
image on the middle row below, we show the same assembly, but the upper ring is
rotated by $\psi=20^o$ and the lower ring by $-20^o$. The images towards the right
of the geometries show the tangential field on the upper panel, the Fourier harmonics
on the middle panel, and the relative magnitude of the unwanted harmonics on the bottom
panel. We see that the dominant harmonic is the quadrupolar $m=2$ with integrated
strength, evaluated at a radius of $5\,$mm, going from $6.26\,$Tmm for $\psi=0^o$
to $4.79\,$Tmm for $\psi=20^o$. The harmonics in all cases are smaller by a few
times $10^{-3}$.
\par
Next, we systematically changed $\psi$ in steps of $5^o$ from $0^o$ to $45^o$ and show
the gradient on the mid-plane (indicate by the black dotted line) on the lower-left
image in Figure~\ref{fig:quad}. We see that the gradient changes from $45.8\,$T/m
for $\psi=0^o$ to zero at $\psi=45^o$. On the right-hand image on the bottom row
we plot the integrated gradient $\int(dB_y/dx)dz$ as a function of $\psi$ in the
upper panel. We see
that it moves smoothly from $6.26\,$Tmm$/5\,$mm$=1.25\,$T for $\psi=0^o$ to zero at
$\psi=45^o$. On the bottom panel we plot the same data versus $\cos2\psi$ and find
a linear dependence of the integrated gradient on $\cos2\psi$, which will make tuning
and setting up double-quadrupole assemblies straightforward.
\section{Some applications}
Here we discuss three examples where the strap-on magnets might prove useful: a small
chicane, a triplet cell, and a solenoid lens. Let's start with the chicane.
\subsection{Chicane}
Four-dipole chicanes are frequently used in accelerators to provide an energy-dependent
path length for a beam, which helps to compress beams longitudinally~\cite{VZAP,BC}.
The figure of merit for a bunch compressor is the so-called $R_{56}\approx -2L\phi^2$, where
$L$ is the distance between the outer and the inner two dipoles, and $\phi$ is the deflection
angle of one dipole. This angle is given by the ratio of the integrated field $\int B_y dz$
of the dipole and the momentum of the beam, often expressed through its rigidity $B\rho$.
For a beam with momentum 300\,MeV/c, the rigidity is about 1\,Tm, such that the variable
two-magnet dipole from Section~\ref{sec:vardip}
with an integrated field of $11.4\,$Tmm deflects a 300\,MeV beam by 11.4\,mrad. With $L=0.4\,$m
between the dipoles we therefore deflect the beam by about 5\,mm and stay within the region
shown on the bottom left in Figure~\ref{fig:dip}. Furthermore, we find
$R_{56}=-0.10\,$mm. Since we can adjust the magnetic field by rotating the two adjacent dipoles
we can adjust the $R_{56}$ between the maximum value and zero.
\par
\begin{figure}[tb]
\begin{center}
\includegraphics[width=0.9\textwidth]{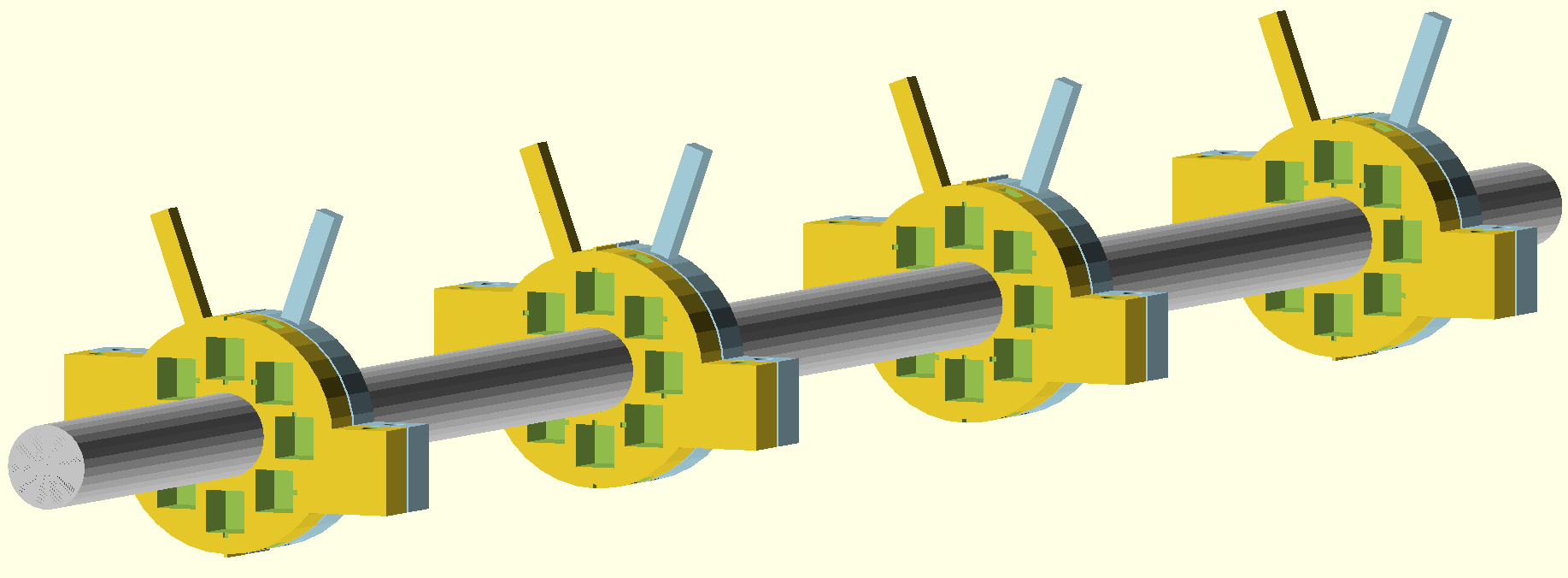}
\end{center}
\caption{\label{fig:chicane}Sketch of the four-dipole chicane made of four double-ring
  variable dipole magnets strapped on a cylindrical pipe.}
\end{figure}
Figure~\ref{fig:chicane} illustrates such a chicane with four variable dipoles wrapped around
the grey beam pipe. One of the magnets is colored light blue and has a handle attached
sticking out to the top right, while the other magnet is colored yellow and has a handle
sticking out towards the top left. The handles are attached to the 5\,mm square holes on
the periphery of the frames that is also visible
on Figure~\ref{fig:frames}. Moreover, careful inspection shows that the two middle
assemblies have opposite polarity (the central ones have a notch sticking out at the top,
while the outer ones have a groove), such that they deflect a beam in opposite directions.
Connecting all yellow handles with a rod, allows to change the rotation angle of all yellow
magnets by the same amount, while a rod connecting the light blue handles rotates all light
blue magnets by the same amount, which permits to change the excitation of all dipoles
synchronously. Small variations in the excitation of different magnets can be compensated by
placing small plates between the handles and the connecting rod. Note that for the
illustration in Figure~\ref{fig:chicane} the distance $L$ was reduced to 0.2\,m.
\subsection{Triplet cell}
%
\begin{figure}[tb]
\begin{center}
\includegraphics[width=0.7\textwidth]{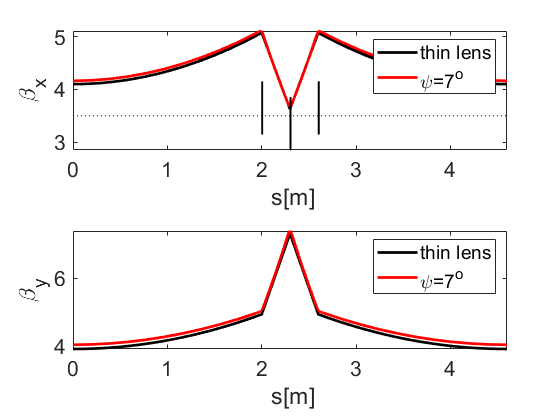}
\end{center}
\caption{\label{fig:tripbeta}The horizontal (top) and vertical (bottom) beta functions
  of a cell made of thin-lens quadrupoles (black lines) with $60^o$ phase advance in
  both planes. The red lines are the corresponding beta functions where the fields of
  quadrupole rings from Figure~\ref{fig:triplet} are used and the central double-ring
  quadrupole is rotated by $\psi=7^o$.}
\end{figure}
In beam lines where long stretches of free space between magnets as well as round beams
are required, often so-called triplets are used. They are combinations of three closely-spaced
quadrupoles where the first and third quadrupole are excited equally and the center quadrupole
has opposite polarity and twice the excitation. Figure~\ref{fig:tripbeta} shows the magnet
lattice with the three thin-lens quadrupoles and the horizontal beta function $\beta_x$ in the
upper panel and the vertical beta function $\beta_y$ in the lower panel. The black lines
show the  beta functions for the thin lens model with a betatron phase advance per cell of $60^o$
in both planes. Note that the magnets are only located over a distance of 0.6\,m such that
there is almost 4\,m free space available between magnets of adjacent cells.
\par
\begin{figure}[tb]
\begin{center}
\includegraphics[width=0.9\textwidth]{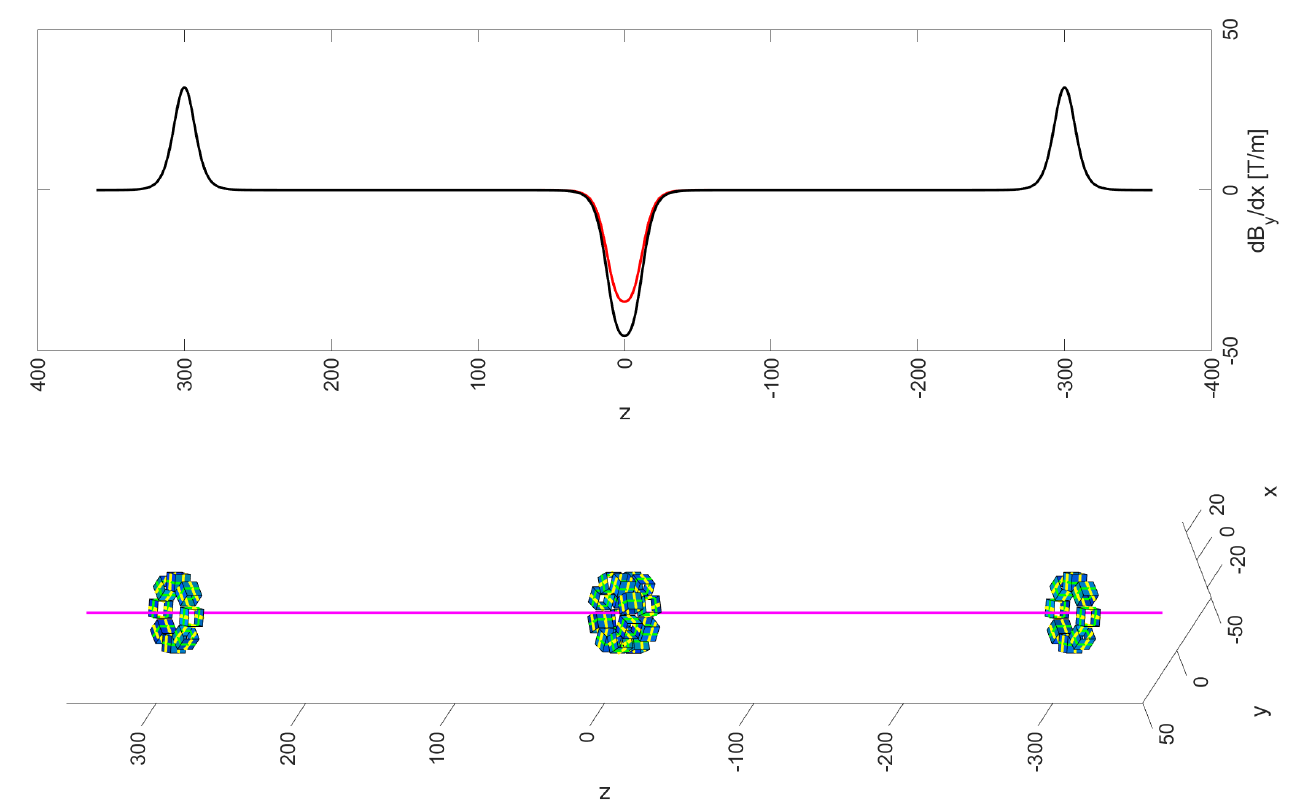}
\end{center}
\caption{\label{fig:triplet}Bottom: four quadrupole rings assembled to form a triplet. The
  strength of the double-ring is adjustable as described in Section~\ref{sec:varquad}. Top:
  the gradient along the beam line for rotation angles of the central quadrupoles of $\psi=0^o$
  (black) and $\psi=20^o$ (red).}
\end{figure}
Our quadrupole rings are particularly attractive to realize such triplets; we use a single
ring for the two outer quadrupoles and two rings, rotated by $90^o$ for the double-strength
center quadrupole with opposite polarity. The lower image in Figure~\ref{fig:triplet} shows
the two outer rings
located near $z=\pm 300\,$mm and the inner double-ring near $z=0\,$mm. The magenta line
denotes the axis on which beam travels. We point out that adjusting the relative rotation
angle $\psi$ between the two center quadrupoles gives us some tunability. The upper
image in Figure~\ref{fig:triplet} shows the gradient along the beam axis. Note the opposite
polarity of the outer and inner quadrupoles. Moreover, the black line shows the field
for $\psi=0^o$ and the red line for $\psi=20^o$.
\par
We then empirically adjust $\psi$, calculate the gradient along the beam axis and
automatically write a beam-line description, consisting of many 1\,mm long quadrupoles,
and calculate the beta function with
software from~\cite{VZAP}. With $\psi=7^o$ we found that the phase advances are close
to the $60^o$ for the thin-lens lattice. The red lines in Figure~\ref{fig:tripbeta}
show $\beta_x$ and $\beta_y$ for the gradients produced by our quadrupole rings, which
agree very well with those of the thin-lens model, shown in black.
\subsection{Solenoid focusing}
%
\begin{figure}[tb]
\begin{center}
\includegraphics[width=0.9\textwidth]{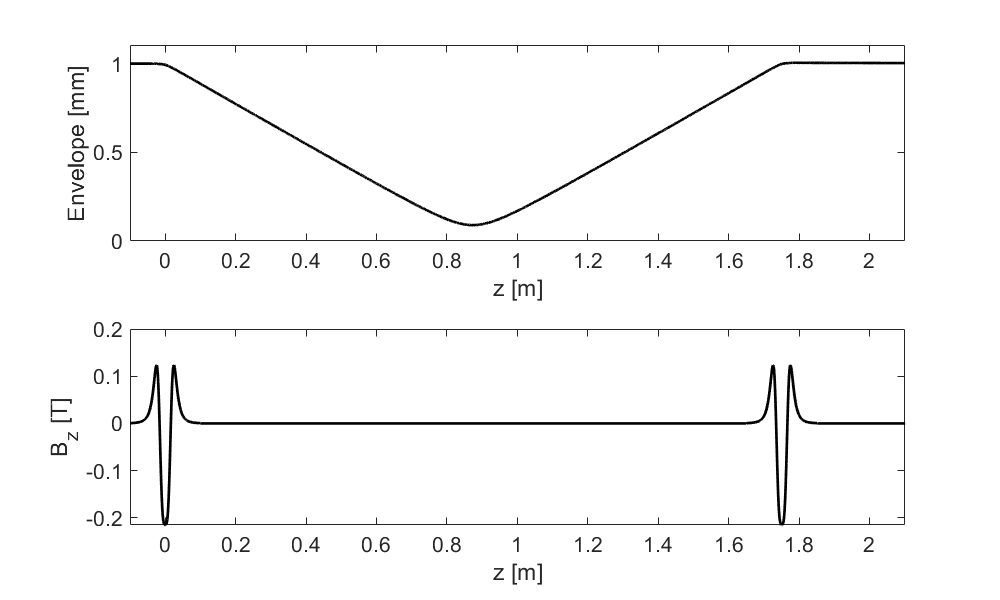}
\end{center}
\caption{\label{fig:solfoc}The longitudinal field $B_z$ created by two radial solenoids
  located at $z=0$ and $z=1.75\,$m (bottom) and the beam envelope of a 5\,MeV electron beam
  with a parallel beam with radius 1\,mm entering from the left (top). We see that the first
  solenoid focuses the beam to a waist near $z=0.87\,$m and the second one makes the beam
  parallel again.}
\end{figure}
The radial solenoid from Section~\ref{sec:solone}, whose geometry and field are shown in
Figure~\ref{fig:solrad}, is suitable to focus round low energy beams. In the upper panel
of Figure~\ref{fig:solfoc} we show the beam envelope of a 5\,MeV electron beam with a
normalized emittance of $\eps_n=10^{-6}\,$m\,rad passing through two such radial solenoids
spaced 1.75\,m apart. On the lower panel we recognize $B_z$ from the right-hand image in
Figure~\ref{fig:solrad}; once near $z=0$ and once near $z=1.75\,$m. The  envelope is calculated
from $B_z$ by numerically integrating the paraxial-ray equation~\cite{REISER}. The initial beam
is parallel and has a radius of 1\,mm, but then develops a focus near $z=0.87\,$m, which is
consistent with the focal length $f$ calculated from $1/f=\int B_z^2dz/(2B\rho)^2\approx 0.87\,$m,
where we have $\int B_z^2dz=1.3\times10^{-3}\,$T$^2$m and $B\rho=0.0169\,$Tm for the 5\,MeV beam.
\par
\begin{figure}[tb]
\begin{center}
\includegraphics[width=0.9\textwidth]{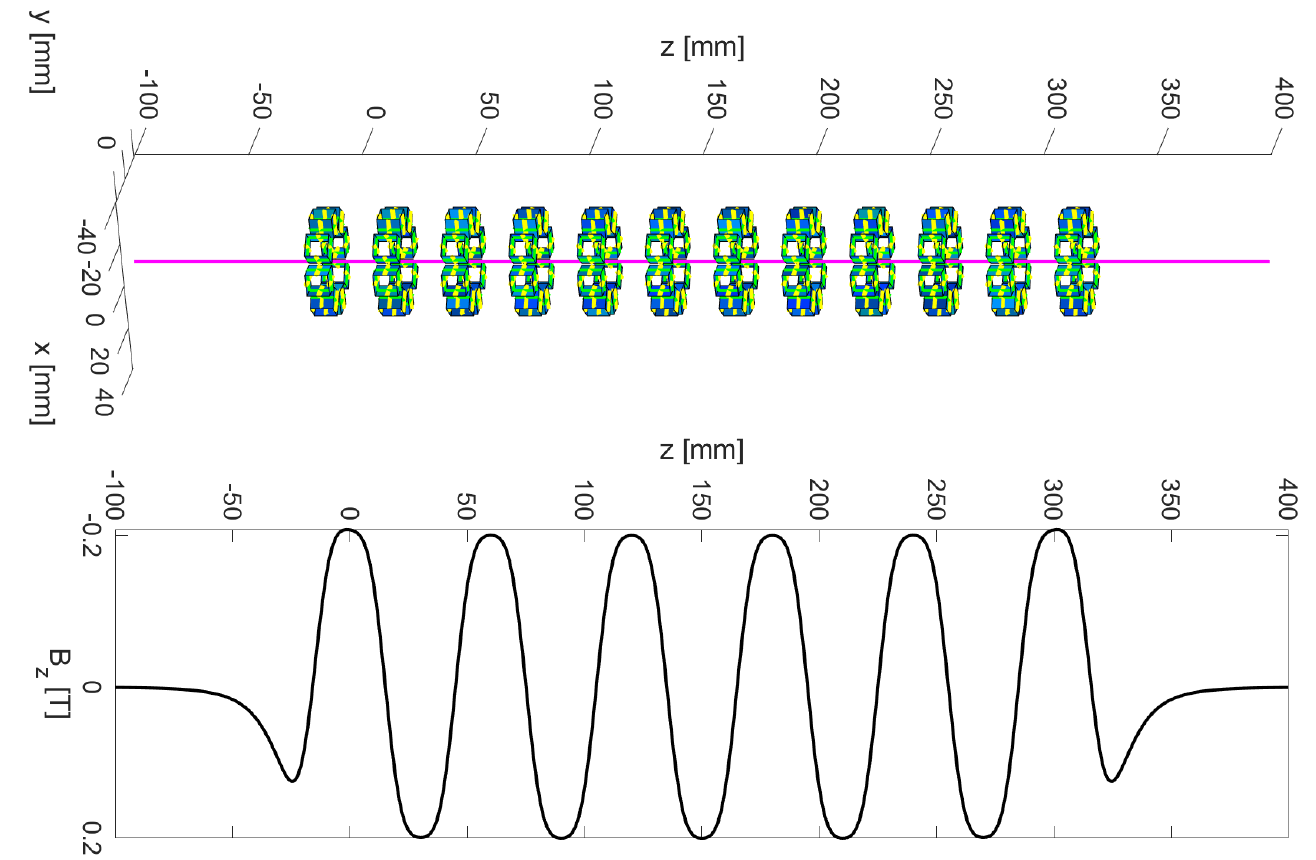}
\end{center}
\caption{\label{fig:solstr}Assembling the six two-ring radial solenoids with alternating polarity,
  shown in the upper panel, creates the longitudinal field $B_z$ shown in the lower panel. It has
  a larger focusing strength than the two ring solenoid used in Figure~\ref{fig:solfoc} and can
  be used for higher energy beams.}
\end{figure}
Making the solenoid stronger is a simple matter of repeating multiple copies of the
radial solenoids, as shown in Figure~\ref{fig:solstr}, where six copies of the solenoid
from Figure~\ref{fig:solrad} follow with equal distance between consecutive rings. Note
that the polarity alternates from one ring to the next. 
The upper panel shows the beam axis in
magenta together with the twelve radial solenoids with alternating polarity, whereas the lower
panel shows the longitudinal component $B_z$ on the beam axis. Note that the rings are
placed every 30\,mm and the amplitude of $B_z$ reaches $\pm0.2\,$T. The integrated
focusing strength $\int B_z^2dz$ is $8.8\times10^{-3}\,$T$^2$m for this configuration. This
configuration would work for a 13\,MeV electron beam ($B\rho\approx0.044\,$Tm) where
it leads to a comparable focal length close to 0.9\,m.
\section{Scaling}
We point out that changing the value of the remanent magnetic field $B_r$ will just scale
the field values reported in the remainder of this report accordingly.
\par
Moreover, changing the linear dimensions, the size of cubes $w$, their radial position $o$,
and the size of the frame, will keep all quantities, whose dimension is Tesla, unchanged.
For example, the field in a dipole made of 5\,mm cubes equals that of a dipole made
of 10\,mm cubes, but its fringe field does not extend as far as that of the bigger dipole.
Therefore the integrated field of the smaller magnet is a factor two smaller. Likewise,
the integrated gradient of a quadrupole made of 5\,mm cubes is the same as that
made of 10\,mm cubes. In the smaller magnet the gradient is doubled, but the total length
is halved.
\section{Conclusions}
Based on the analytic results for the magnetic fields from permanent magnets
from~\cite{VZCFE} we explored the use of permanent-magnet cubes to create
dipoles, quadru\-poles, and solenoids. All examples are based on the frames
for 10\,mm cubes, shown in Figure~\ref{fig:frames}, but are easily scalable
to use cubes of a different size, to include more cubes, or produce other
multipoles. We found that the magnets have reasonably good field qualities; the
amplitudes of the harmonics at a radius of 5\,mm in all cases are less than
a fraction of a percent and often better.
\par
An attractive feature is the possibility to retrofit these magnets to existing
beam lines without having to open the vacuum system; the two magnet halves are
simply
strapped-on to the beam pipe. A second attractive feature is the possibility
to combined two closely spaced rings to obtain dipoles and quadrupoles whose
strength is continuously adjustable, while their field quality is comparable to
that of a single ring.
\par
Finally, we found these magnet rings suitable to construct beam-line modules
for chicanes, a triplet cell, and solenoids focusing a low-energy beam.
\par
\subsection*{Acknowledgments}
In part, funding is
provided through the project {\em Disseminating technology for cold magnets to
  provide access to a wider international market} that is supported by the
European Regional Development Fund (ERDF) and Region Kronoberg.

%
%
\bibliographystyle{plain}

\begin{thebibliography}{M}
  %
\bibitem{KH1}
  K. Halbach, {\em Design of permanent multipole magnets with oriented
    rare earth cobald magnets,} Nuclear Instruments and Methods 169 (1980) 1.
\bibitem{VZAP}
  V. Ziemann, {\em Hands-On Accelerator Physics Using MATLAB,} CRC Press, Boca Raton, 2019.
\bibitem{VZCFE}
  V. Ziemann, {\em Closed-form expressions for the magnetic field of permanent
    magnets in three dimensions,} arXiv:2106.04153, June 2021.
\bibitem{BC}
  F. Stulle, A. Adelmann, M. Pedrozzi, {\em Designing a bunch compressor chicane
    for a multi-TeV linear collider,}  Phys.Rev.ST Accel.Beams 10 (2007) 031001.
\bibitem{REISER}
  M. Reiser, {\em Theory and design of charged particle beams,} Wiley-VCH, Weinheim, 2004.
%
\end{thebibliography}

\end{document}